\documentclass[journal]{IEEEtran}
\ifCLASSINFOpdf
\else
   \usepackage{graphicx}
   \graphicspath{{../eps/}}
   \DeclareGraphicsExtensions{.eps}
\fi

\usepackage{amsmath}
\usepackage{amsfonts,amssymb}
\usepackage{amssymb}
\usepackage{wrapfig}
\usepackage{psfrag}
\usepackage{epstopdf}
\usepackage{cite}
\usepackage{graphicx}
\usepackage{subfigure}
\usepackage{threeparttable}
\usepackage{cases}
\usepackage{subeqnarray}
\usepackage{color}
\usepackage{underscore}
\usepackage{verbatim}
\usepackage{bm}
\usepackage{stfloats}
\usepackage{xcolor}
\usepackage{xpatch}

\newtheorem{remark}{Remark}

\newtheorem{algorithm}{Algorithm}
\newtheorem{proposition}{Proposition}

\usepackage{algorithm}
\usepackage{algorithmic}

\begin{document}
\title{Aerial Intelligent Reflecting Surface: \\Joint Placement and Passive Beamforming \\Design with 3D Beam Flattening }

\author{
        Haiquan~Lu,
        Yong~Zeng,~\IEEEmembership{Member,~IEEE,}
        Shi~Jin,~\IEEEmembership{Senior Member,~IEEE,}
        and
        Rui~Zhang,~\IEEEmembership{Fellow,~IEEE}
\thanks{This work was supported by the National Key R\&D Program of China with Grant number 2019YFB1803400. Part of this work has been presented at the IEEE ICC 2020 Workshop, Dublin, Ireland, 7-11 June 2020 ~\cite{lu2020enabling}.}
\thanks{H. Lu, Y. Zeng, and S. Jin are with the National Mobile Communications Research Laboratory, Southeast University, Nanjing 210096, China, Y. Zeng is also with the Purple Mountain Laboratories, Nanjing 211111, China (e-mail: \{haiquanlu, yong_zeng, jinshi\}@seu.edu.cn). (\emph{Corresponding author: Yong Zeng.})}

\thanks{R. Zhang is with the Department of Electrical and Computer Engineering, National University of Singapore, Singapore 117583 (e-mail: elezhang@nus.edu.sg).}
}

\maketitle


\begin{abstract}
  Intelligent reflecting surface (IRS) is a promising technology to reconfigure wireless channels, which brings a new degree of freedom for the design of future wireless networks. This paper proposes a new three-dimensional (3D) wireless passive relaying system enabled by \emph{aerial IRS} (AIRS). Compared to the conventional terrestrial IRS, AIRS enjoys more deployment flexibility as well as wider-range signal reflection, thanks to its high altitude and thus more likelihood of establishing line-of-sight (LoS) links with ground source/destination nodes. Specifically, we aim to maximize the worst-case signal-to-noise ratio (SNR) over all locations in a target area by jointly optimizing the transmit beamforming for the source node and the placement as well as 3D passive beamforming for the AIRS. The formulated problem is non-convex and thus difficult to solve. To gain useful insights, we first consider the special case of maximizing the SNR at a given target location, for which the optimal solution is obtained in closed-form. The result shows that the optimal horizontal AIRS placement only depends on the ratio between the source-destination distance and the AIRS altitude. Then for the general case of AIRS-enabled area coverage, we propose an efficient solution by decoupling the AIRS passive beamforming design to maximize the worst-case \emph{array gain}, from its placement optimization by balancing the resulting \emph{angular span} and the cascaded channel path loss. Our proposed solution is based on a novel \emph{3D beam broadening and flattening} technique, where the passive array of the AIRS is divided into sub-arrays of appropriate size, and their phase shifts are designed to form a flattened beam pattern with adjustable beamwidth catering to the size of the coverage area. Both the uniform linear array (ULA)-based and uniform planar array (UPA)-based AIRSs are considered in our design, which enable two-dimensional (2D) and 3D passive beamforming, respectively. Numerical results show that the proposed designs achieve significant performance gains over the benchmark schemes.
\end{abstract}


\IEEEpeerreviewmaketitle
\section{Introduction}
While the fifth-generation (5G) wireless communication network is being deployed worldwide, research on the next/sixth-generation (6G) wireless network has embarked. As a key driver for the future intelligent information empowered society, 6G is expected to provide pervasive connectivity with data rate 100-1000 times higher than that of 5G, i.e., up to 1 Tera-byte per second (Tbps)~\cite{Latvaaho2019Keydrivers,letaief2019roadmap}. To this end, several key wireless transmission technologies, such as Terahertz communication and ultra-massive multiple-input multiple-output (UM-MIMO) have received significant research attention \cite{yang20196g}. Despite of the great potential for drastic performance improvement by such technologies, their required large antenna arrays at high carrier frequency render practical implementation issues such as hardware cost, power consumption and signal processing complexity more severe. Therefore, developing high-capacity yet cost-effective communication techniques is of paramount importance for 6G.

During the past years, various cost-effective wireless communication techniques have been proposed at the transmitter and/or receiver side, such as analog beamforming~\cite{wang2009beam}, hybrid analog/digital beamforming~\cite{zhang2005variable-phase-shift-based,ayach2014spatially}, lens MIMO communications~\cite{zeng2016millimeter}, and low-resolution analog-to-digital converters (ADCs)~\cite{singh2009on,zhang2018on}. More recently, wireless communication aided by intelligent reflecting surface (IRS) has emerged as a new promising technique for achieving cost-effective wireless communications via proactively manipulating the radio environment~\cite{subrt2012intelligent,tan2016increasing,tang2019wireless,di2019smart,han2019large,huang2019reconfigurable,
basar2019wireless,cui2019secure,wu2019intelligent,wu2019towards,bjornson2020intelligent,Yang2020IntelligentRS,
Tang2019WirelessCW,wu2020intelligentTutorial,huang2020achievable}. IRS is a man-made reconfigurable metasurface composed of a large number of regularly arranged sub-wavelength passive elements and a smart controller \cite{wu2019intelligent,wu2019towards}. Through modifying the amplitude and/or phase of the impinging radio waves, IRS is able to dynamically control the radio propagation for various purposes, such as signal enhancement, interference suppression and transmission security\cite{wu2019towards}. Different from the conventional active relays, the radio signal reflected by IRS is free from self-interference or noise corruption in an inherently full-duplex manner. IRS-aided wireless communication has been studied from different aspects, such as energy efficiency maximization~\cite{huang2019reconfigurable}, secrecy rate maximization~\cite{cui2019secure}, joint active and passive beamforming design~\cite{wu2019intelligent}, and rate region characterization for IRS-aided interference channel~\cite{huang2020achievable}, etc. Besides the aforementioned passive reflecting metasurface, active components such as radio frequency (RF) circuits and signal processing units can also be embedded in the metasurface to serve as transceivers, known as active surfaces, which may drastically reduce the hardware complexity and energy cost compared to conventional antenna arrays~\cite{huang2020holographic,alexandropoulos2020reconfigurable,shlezinger2020dynamic}.

\begin{figure}[!t]
  \centering
  \subfigure[Terrestrial IRS]{
  \hspace{8mm}
    \includegraphics[width=1.2in,height=0.94in]{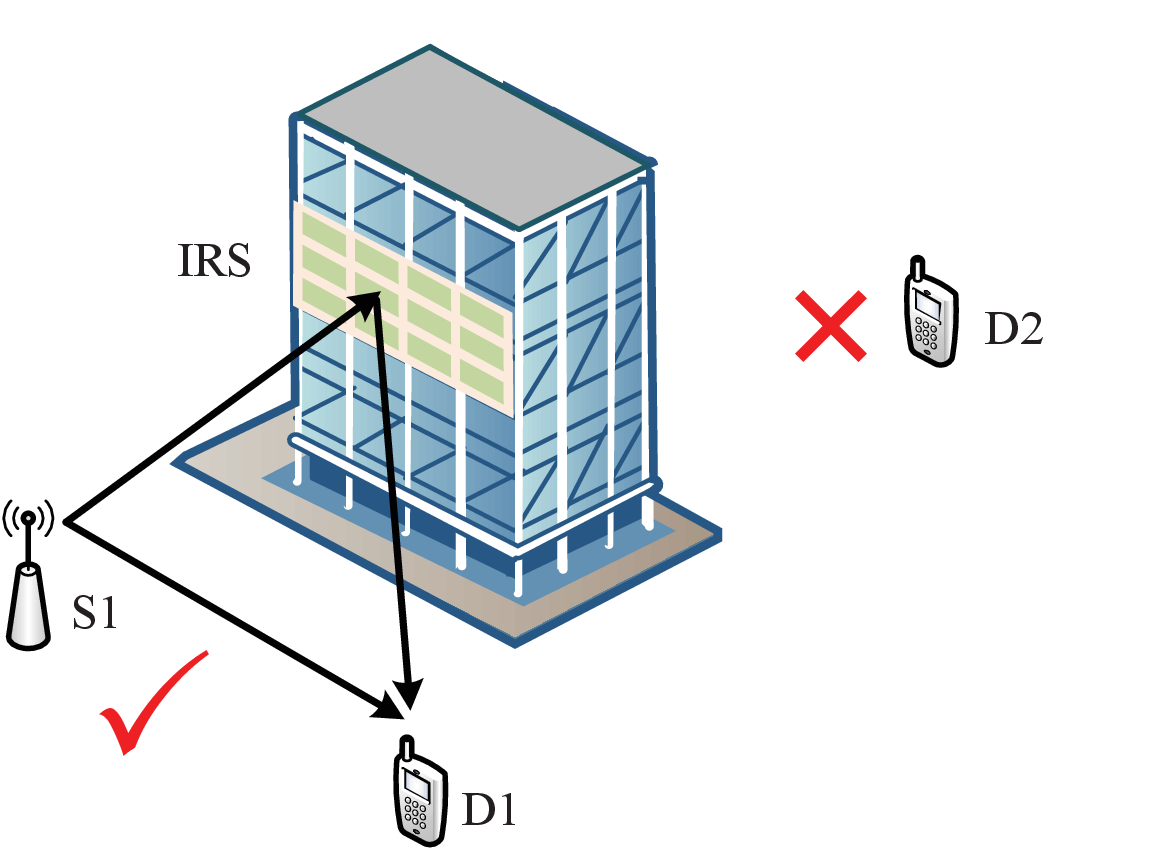}
  }
  \subfigure[AIRS]{
    \includegraphics[width=1.2in,height=1.28in]{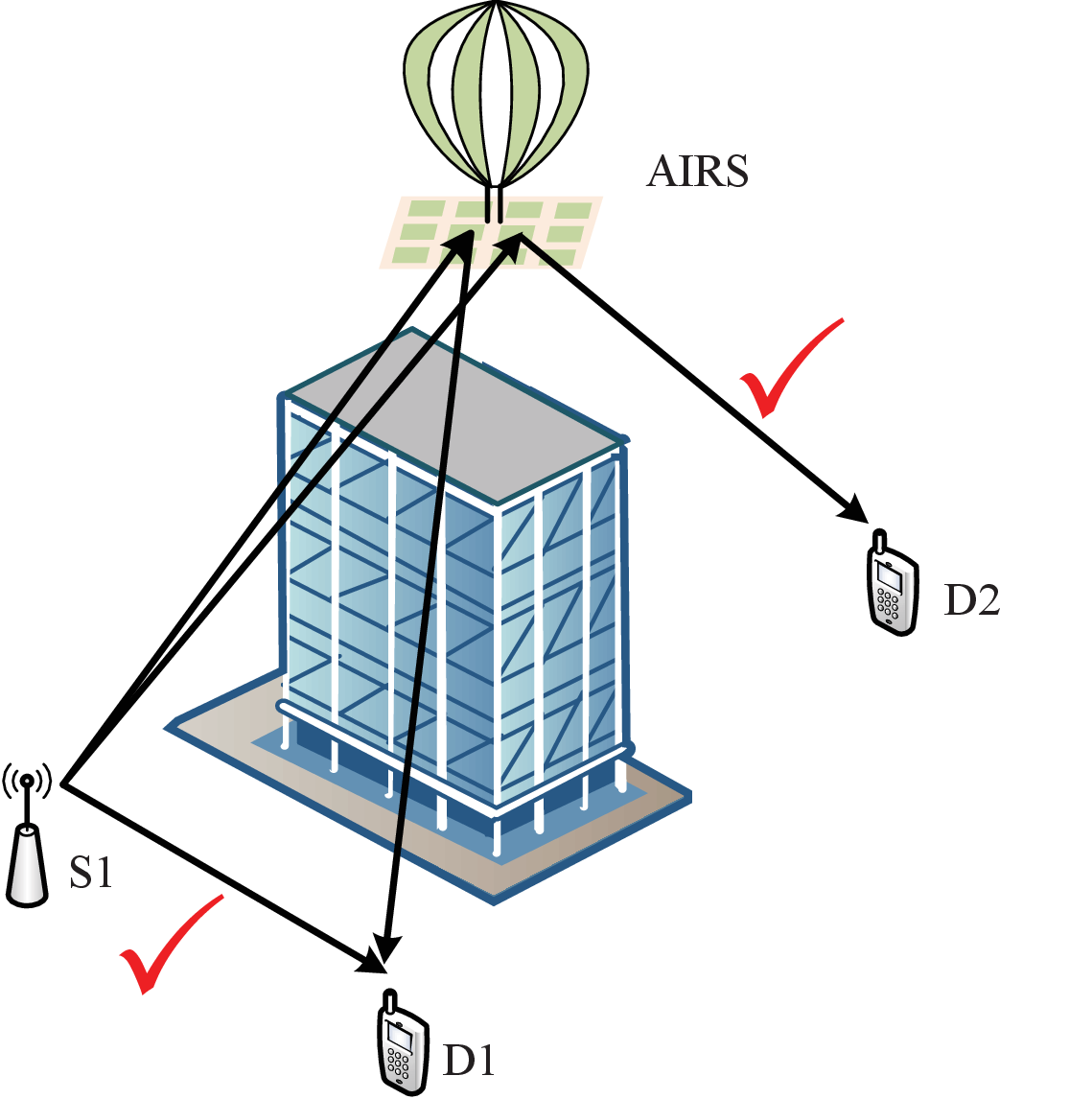}
  }
  \caption{Half-space reflection by terrestrial IRS versus panoramic/full-angle reflection by AIRS.}
  \label{half and full of the space}
  \end{figure}

\begin{figure}[!t]
  \centering
  \subfigure[Terrestrial IRS]{
    \includegraphics[width=1.2in,height=1.1in]{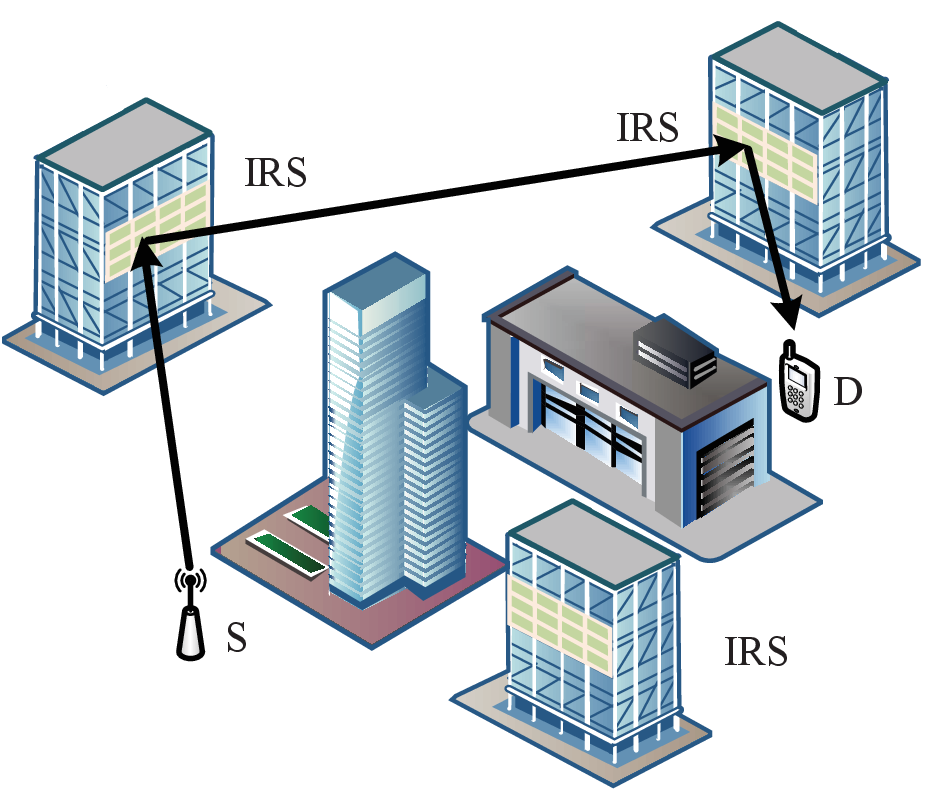}
  }
  \hspace{5mm}
  \subfigure[AIRS]{
    \includegraphics[width=1.2in,height=1.35in]{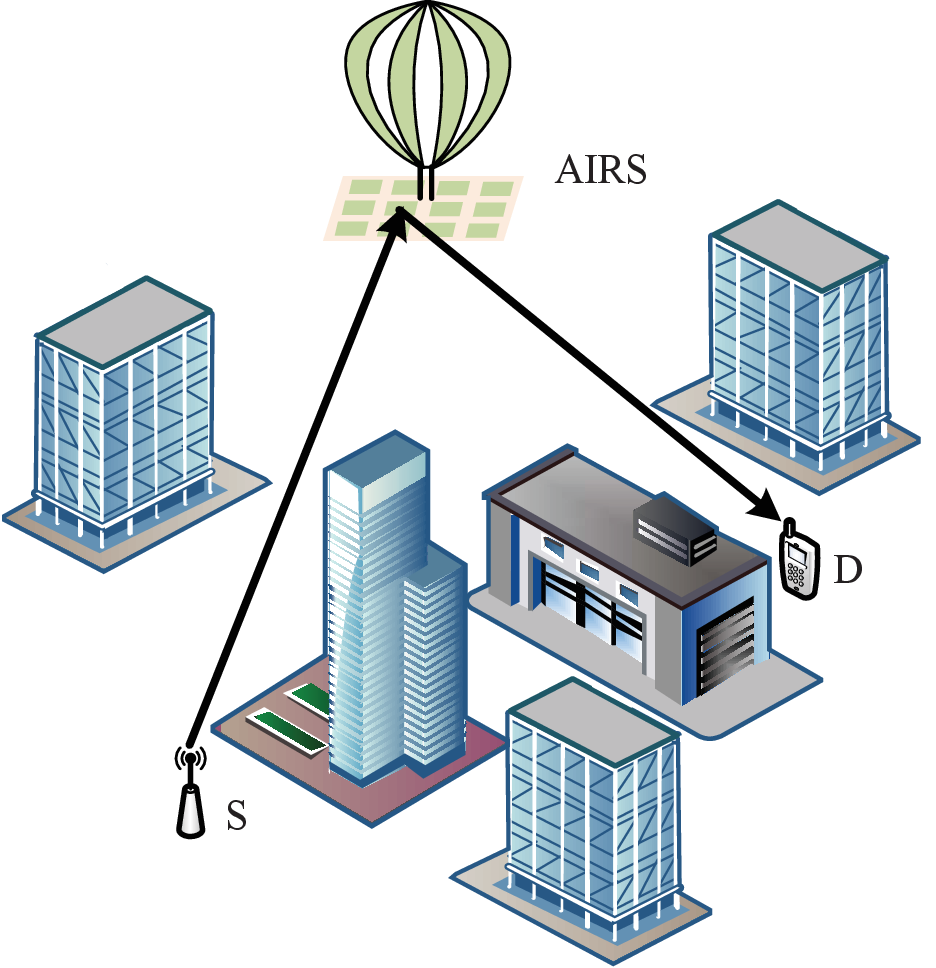}
  }
  \caption{AIRS can reduce the number of reflections than terrestrial IRS.}
  \label{Different reflection times between IRS deployed on the facade of building and aerial platforms}
  \end{figure}

However, most existing research on IRS-aided communication focuses on terrestrial IRS that is deployed on e.g., facades of buildings or indoor walls/ceilings. Such an IRS architecture poses fundamental limitations for several reasons. First, from the deployment perspective, finding the appropriate place for IRS installation is usually difficult in practice, since it involves various issues like site rent, impact of urban landscape and the willingness of owners to install large IRS on their properties. Second, from the performance perspective, IRS deployed on the walls or facades of buildings can at most serve terminals located in half of the space, i.e., both the source and destination nodes must lie on the same side of the IRS, as illustrated in Fig.~\ref{half and full of the space}(a). Third, as shown in Fig.~\ref{Different reflection times between IRS deployed on the facade of building and aerial platforms}(a), in complex environment like urban areas, the radio signals originated from a source node typically have to undergo several reflections before reaching the desired destination. This thus leads to significant signal attenuation since each reflection, even with IRS-enabled passive beamforming, would still cause signal scattering to undesired directions.

To address the above issues, we propose in this paper a novel three-dimensional (3D) wireless network enabled by aerial IRS (AIRS), where IRS is mounted on aerial platforms like balloon, unmanned aerial vehicle (UAV), so as to enable intelligent reflection from the sky. Compared to the conventional terrestrial IRS, AIRS has several appealing advantages. First, with elevated position, AIRS can more easily establish line-of-sight (LoS) links with the ground nodes~\cite{zeng2019accessing}, which leads to stronger channel as compared to the terrestrial IRS. At the same time, the placement or trajectory of aerial platforms can be more flexibly optimized to further improve the communication performance, thereby offering a new  degree of freedom (DoF) for performance enhancement via 3D network design. Second, AIRS is able to achieve panoramic/full-angle reflection, i.e., one AIRS can in principle help reflect signals between any pair of nodes on the ground, as illustrated in Fig.~\ref{half and full of the space}(b). This is in a sharp contrast to the conventional terrestrial IRS that can only serve nodes in half of the space. Last but not least, in contrast to the terrestrial IRS, AIRS is usually able to achieve desired signal manipulation by one reflection only, even in complex urban environment (see Fig.~\ref{Different reflection times between IRS deployed on the facade of building and aerial platforms}(b)), thanks to its high likelihood of having LoS links with the ground nodes. This thus greatly reduces the signal power loss due to multiple reflections with the terrestrial IRS.

Despite of many advantages mentioned above, AIRS faces several new challenges, such as the endurance, stability and controllability of the aerial platform carrying the IRS. The additional flight control unit and safety measures required to deploy or fly AIRS, the impact of drift or vibration of the aerial platform on the performance and design of AIRS, as well as the challenges in channel estimation for AIRS-aided communication all deserve further investigation.

In this paper, we consider a basic setup of AIRS-enabled wireless relaying system, where an AIRS is deployed to help extend the signal coverage from a ground source node (e.g., base station (BS)/access point (AP)) to a given target area, say, a hot spot in cellular network or a remote area without cellular coverage. Our objective is to maximize the worst-case/minimum signal-to-noise ratio (SNR) in the target area by jointly optimizing the transmit beamforming for the source node and the placement as well as 3D passive beamforming for the AIRS. The formulated optimization problem is difficult to be directly solved due to the following reasons. First, different from most of the existing research where the passive beamforming of the terrestrial IRS is designed based on the channel state information (CSI) of users at known locations, the beamforming optimization for the AIRS needs to balance the received SNRs at all locations in the target area, which results in a more complicated design problem. Second, different from terrestrial IRS deployed at fixed location, the AIRS placement is a new problem to solve, which affects not only the source-AIRS-destination cascaded channel path loss, but also the angle of departures (AoDs) and angle of arrivals (AoAs) of the source-AIRS link as well as the \emph{angular span} from the AIRS to the target area. The main contributions of this paper are summarized as follows.

First, for the general uniform planar array (UPA)-based AIRS, an optimization problem is formulated to maximize the worst-case SNR over all locations in the target area by jointly optimizing the transmit beamforming at the source node and the AIRS placement as well as 3D passive beamforming.{\footnote[1]{This facilitates the multiple access design for users at random locations in the target area, via e.g., time-division multiple-access (TDMA) or frequency-division multiple-access (FDMA).}} We show that the optimal transmit beamforming for the source node corresponds to the well-known maximum ratio transmission (MRT). Therefore, the problem is reduced to the joint optimization of the AIRS placement and 3D passive beamforming for the min-SNR maximization in the target area.

Next, we consider the special case of SNR maximization at a given target location, for which the optimal AIRS placement and phase shifts for passive beamforming are derived in closed-form. The solution shows that the optimal horizontal AIRS placement depends on $\rho$, which is the ratio between the source-destination distance and the AIRS altitude. For $0 \le \rho  \le 2$, the AIRS should always be placed above the midpoint between the source node and the destination location. On the other hand, when $\rho  > 2$, there exist two optimal horizontal AIRS placement locations that are symmetric over the midpoint.

Last, for the general case of min-SNR maximization in a target area, we propose an efficient two-step solution by decoupling the AIRS passive beamforming design and its placement optimization. The key of the proposed solution lies in a novel \emph{beam broadening and flattening} technique, where the passive array is partitioned into multiple sub-arrays with their phase shifts optimized to form one single flattened beam with its beamwidth properly tuned to match the size of the target coverage area. For ease of exposition, we first consider the special case of uniform linear array (ULA)-based AIRS, for which passive beam flattening only needs to be applied in one spatial frequency dimension. Then the proposed design is extended to the general UPA-based AIRS, which enables 3D passive beam flattening over two spatial frequency dimensions. By leveraging the proposed beam broadening and flattening technique, the optimization of AIRS placement is greatly simplified, which only needs to balance the resulting angular span and the cascaded path loss.

Notice that wireless communication aided by aerial platforms (e.g., balloons and UAVs) has received significant attention recently (see \cite{zeng2019accessing} and references therein). However, most of such existing works are based on the conventional active communication techniques, such as active aerial signal transmission/receiving/relaying. In fact, the appealing advantages of the passive IRS, such as its compact size, light weight, low energy consumption, and conformal geometry, make it a promising alternative for aerial platforms than conventional mobile relays based on active communication techniques. On the other hand, beam broadening/flattening technique has received increasing attention recently, mainly in the context of analog beamforming for cost-effective designs at the transmitter side~\cite{wang2009beam,hur2013millimeter,zhang2017codebook,zhu20193D}. Different from the existing work, in this paper, we provide a systematic and rigorous derivation for the 3D passive beam broadening/flattening for AIRS-aided communications, so as to form a 3D passive beam pattern that matches the size of the target coverage area with nearly equal gains over all locations therein.{\footnote[2]{This technique can be extended to the analog beamforming design of active metasurfaces (e.g., dynamic metasurface antennas (DMAs) in~\cite{shlezinger2020dynamic}). }}  The derived results greatly simplifies the subsequent AIRS placement optimization.

The rest of this paper is organized as follows. Section II introduces the system model and the problem formulation for the AIRS-enabled wireless relaying system. In Section III and IV, we propose efficient algorithms to solve the formulated problems in the single target location and area coverage cases, respectively. Section V presents numerical results to evaluate the performance of the proposed designs. Finally, we conclude the paper in Section VI.

\emph{Notations:} Scalars are denoted by italic letters. Vectors and matrices are denoted by bold-face lower- and upper-case letters, respectively. ${{\mathbb{C}}^{M \times N}}$ denotes the space of $M \times N$ complex-valued matrices. For a vector $\bf{x}$, $\left\| {\bf{x}} \right\|$ denotes its Euclidean norm. $\rm{diag}({\bf{x}})$ denotes a diagonal matrix with its diagonal elements given by $\bf{x}$. The symbol $j$ denotes the imaginary unit of complex numbers, with ${j^2} =  - 1$. For a real number $x$, $\left\lceil x \right\rceil$ denotes the smallest integer that is greater than or equal to $x$. The symbol $ \otimes $ denotes the Kronecker product operation.
\section{System Model And Problem Formulation}
 \begin{figure}[!t]
  \centering
  \centerline{\includegraphics[width=3.5in,height=2.0in]{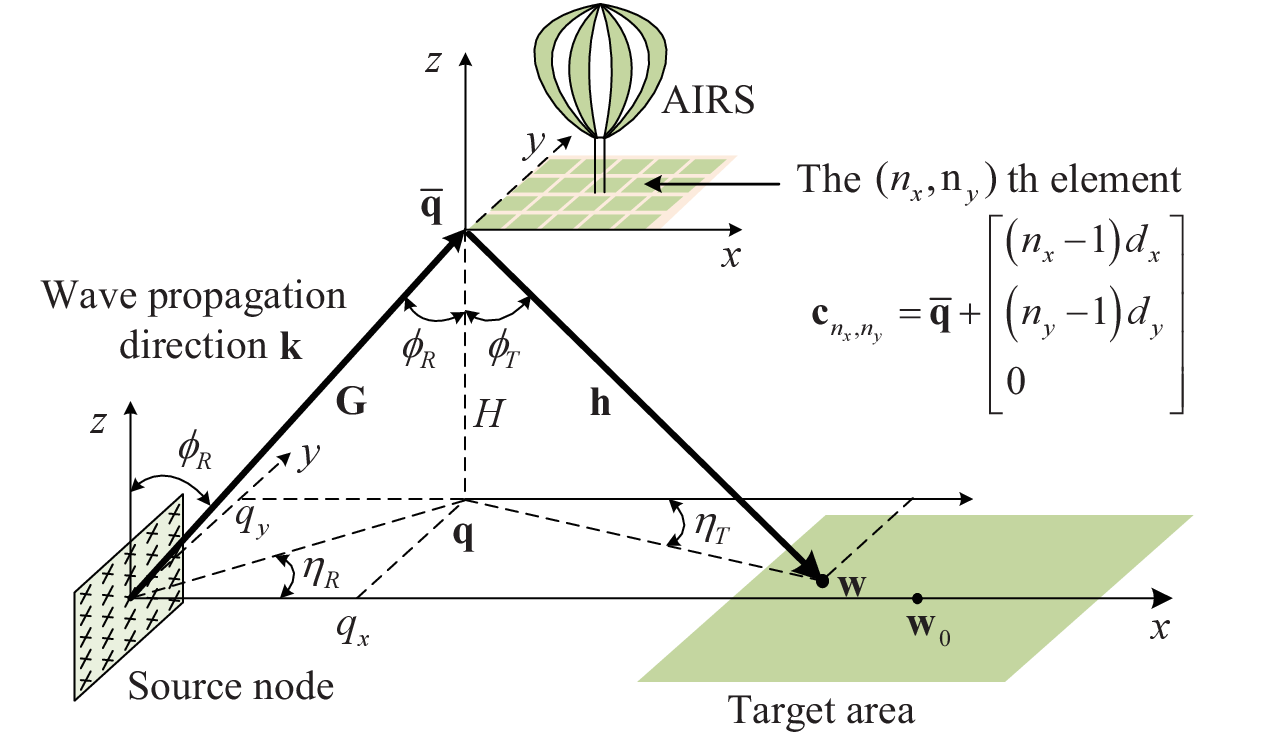}}
  \caption{AIRS-enabled wireless communication system.}
  \label{systemmodel}
  \end{figure}
  As shown in Fig.~\ref{systemmodel}, we consider an AIRS-enabled wireless relaying system, where IRS mounted on an aerial platform (e.g., balloon or UAV) is deployed to assist a terrestrial source node (e.g., BS/AP) to extend its communication coverage to a given terrestrial area of interest, $\mathcal A$. We assume that the direct link from the source node to the target area is negligible due to severe terrestrial blockage/shadowing. Without loss of generality, we assume that the source node is located at the origin of a 3D Cartesian coordinate system and the center of the target area is on the $x$-axis, which is denoted as ${{\bf{w}}_0} = {\left[ {{x_0},0} \right]^T}$ on the $x$-$y$ plane. For ease of exposition, we assume that $\mathcal A $ is a rectangular area on the $x$-$y$ plane. Therefore, any location in $\mathcal A$ can be specified as ${\bf{w}} = {\left[ {{w_x},{w_y}} \right]^T}$ on the $x$-$y$ plane, ${w_x} \in \left[ {{x_0} - \frac{{{D_x}}}{2},{x_0} + \frac{{{D_x}}}{2}} \right], {w_y} \in \left[ { - \frac{{{D_y}}}{2},\frac{{{D_y}}}{2}} \right]$, with $D_x$  and $D_y$ denoting the length and width of the rectangular area, respectively. The AIRS consists of a sub-wavelength UPA with $N = {N_x}{N_y}$ passive reflecting elements, where $N_x$ and $N_y$ denote the number of elements along the $x$- and $y$-axis, respectively. The adjacent elements are separated by ${d_x} < \frac{\lambda }{2}$ and ${d_y} < \frac{\lambda }{2}$, respectively, where $\lambda$ denotes the signal wavelength. The source node is assumed to be equipped with a conventional UPA placed on the $y$-$z$ plane, and the number of antennas is $M={M_y}{M_z}$ with $M_y$ and $M_z$ denoting the number of elements along the $y$- and $z$-axis, respectively.

  We assume that the altitude of the AIRS is fixed at $H > 0$. Without loss of generality, as shown in Fig.~\ref{systemmodel}, we take the bottom-left element of the AIRS as the reference point to represent the AIRS's horizontal location, whose coordinate is denoted by ${{\bf{q}}} = {\left[ {{q_x},{q_y}} \right]^T}$ on the $x$-$y$ plane.  Therefore, the distance from the source node to the AIRS, and that from the AIRS to any location of the target area ${\bf{w}} \in \mathcal A$ can be expressed as {\small{${d_{\bf{G}}} = \sqrt {{H^2} + {{\left\| {{{\bf{q}}}} \right\|}^2}} $}} and {\small{${d_{\bf{h}}} = \sqrt {{H^2} + {{\left\| {{{\bf{q}}} - {\bf{w}}} \right\|}^2}} $}}, respectively.

  In practice, the communication links between the aerial platform (with sufficiently large $H$) and ground nodes are dominated by LoS with high probability~\cite{zeng2019accessing}. Thus, for simplicity, we assume that the aerial-ground channel follows the free-space path loss model, and the channel power gain from the source node to the AIRS can be expressed as
  \begin{equation}
  {\beta _{\bf{G}}}\left( {\bf{q}} \right) = {\beta _0}d_{\bf{G}}^{ - 2} = \frac{{{\beta _0}}}{{{H^2} + {{\left\| {\bf{q}} \right\|}^2}}},
  \end{equation}
  where ${\beta _{\rm{0}}}$ represents the channel power at the reference distance ${d_0} = 1$~m. Similarly, the channel power gain from the AIRS to any location $\mathbf w\in \mathcal A$ can be expressed as
  \begin{equation}
  {\beta _{\bf{h}}}\left( {{\bf{q}},{\bf{w}}} \right) = {\beta _0}d_{\bf{h}}^{ - 2} = \frac{{{\beta _0}}}{{{H^2} + {{\left\| {{\bf{q}} - {\bf{w}}} \right\|}^2}}}.
  \end{equation}

  Note that in practice, the AIRS size is much smaller than the link distances between the AIRS and source/destination nodes. Therefore, the signal from the source node to the AIRS and that from the AIRS to the destination node can be well approximated as uniform plane waves. As illustrated in Fig.~\ref{systemmodel}, denote by ${{\phi _R}\left( {\bf{q}} \right)}$ the zenith AoA of the signal from the source node to the AIRS, i.e., the angle between the wave propagation direction and the positive $z$-axis, and ${\eta _R}\left( {\bf{q}} \right)$ the azimuth AoA, i.e., the angle between the horizontal projection of the wave propagation direction and the positive $x$-axis. The receive array response of the AIRS, denoted as ${{\bf{a}}_R}\left( {{\phi _R}\left( {\bf{q}} \right),{\eta _R}\left( {\bf{q}} \right)} \right)$, is thus dependent on the AIRS (horizontal) placement $\bf{q}$, which is derived as follows. As shown in Fig.~\ref{systemmodel}, with the AIRS located at ${\bf{\bar q}} = {\left[ {{q_x},{q_y},H} \right]^T}$ in 3D, the wave propagation direction of the signal from the source node to the AIRS is ${\bf{k}} = \frac{{{\bf{\bar q}}}}{{\left\| {{\bf{\bar q}}} \right\|}}$, and the coordinate of the $\left( {{n_x},{n_y}} \right)$th AIRS element is ${{\bf{c}}_{{n_x},{n_y}}} = {\bf{\bar q}} + {[\left( {{n_x} - 1} \right){d_x},\left( {{n_y} - 1} \right){d_y},0]^T}, 1 \le {n_x} \le {N_x}, 1 \le {n_y} \le {N_y}$. Then the phase delay of the $\left( {{n_x},{n_y}} \right)$th element relative to the reference element at ${{\bf{\bar q}}}$ is
\begin{equation}
\begin{aligned}
{\psi_{{n_x},{n_y}}} &= \frac{{2\pi }}{\lambda }{{\bf{k}}^T}\left( {{{\bf{c}}_{{n_x},{n_y}}} - {\bf{\bar q}}} \right) = \frac{{2\pi }}{\lambda }\frac{{{{{\bf{\bar q}}}^T}}}{{\left\| {{\bf{\bar q}}} \right\|}}\left[ \begin{array}{l}
\left( {{n_x} - 1} \right){d_x}\\
\left( {{n_y} - 1} \right){d_y}\\
0
\end{array} \right]\\
 &= \frac{{2\pi }}{\lambda }\left( {\frac{{{q_x}}}{{\left\| {{\bf{\bar q}}} \right\|}}\left( {{n_x} - 1} \right){d_x} + \frac{{{q_y}}}{{\left\| {{\bf{\bar q}}} \right\|}}\left( {{n_y} - 1} \right){d_y}} \right). \label{phasedelay1}
\end{aligned}
\end{equation}
Furthermore, it is observed from Fig.~\ref{systemmodel} that the horizontal coordinates of the AIRS can be expressed as ${q_x} = \left\| {{\bf{\bar q}}} \right\|\sin \left( {{\phi _R}} \right)\cos \left( {{\eta _R}} \right)$ and ${q_y} = \left\| {{\bf{\bar q}}} \right\|\sin \left( {{\phi _R}} \right)\sin \left( {{\eta _R}} \right)$. We thus have
\begin{equation}
\begin{aligned}
{\psi _{{n_x},{n_y}}} = \frac{{2\pi }}{\lambda }\left( {\left( {{n_x} - 1} \right){d_x}\sin \left( {{\phi _R}} \right)\cos \left( {{\eta _R}} \right) + } \right.\\
\left. {\left( {{n_y} - 1} \right){d_y}\sin \left( {{\phi _R}} \right)\sin \left( {{\eta _R}} \right)} \right).\label{phasedelay2}
\end{aligned}
\end{equation}
Thus, the corresponding complex coefficients for the $\left( {{n_x},{n_y}} \right)$th element is ${e^{ - j{\psi _{{n_x},{n_y}}}}},1 \le {n_x} \le {N_x},1 \le {n_y} \le {N_y}$. By concatenating such complex coefficients of all the $N = {N_x}{N_y}$ elements, the receive array response vector of the AIRS can be expressed as
\begin{equation}
\begin{aligned}
&{{\bf{a}}_R}\left( {{\phi _R}\left( {\bf{q}} \right),{\eta _R}\left( {\bf{q}} \right)} \right) = \\
&{\left[ {1, \cdots ,{e^{ - j2\pi \left( {{N_x} - 1} \right){{\bar d}_x}{{\bar \Phi }_R}\left( {\bf{q}} \right)}}} \right]^T} \otimes\\
&{\left[ {1, \cdots ,{e^{ - j2\pi \left( {{N_y} - 1} \right){{\bar d}_y}{{\bar \Omega }_R}\left( {\bf{q}} \right)}}} \right]^T},\label{UPAReceiveArrayResponseAtAIRS}
\end{aligned}
\end{equation}
where ${{\bar d}_x} \buildrel \Delta \over = \frac{{{d_x}}}{\lambda }$, ${{\bar d}_y} \buildrel \Delta \over = \frac{{{d_y}}}{\lambda }$, ${{\bar \Phi }_R}\left( {\bf{q}} \right) \buildrel \Delta \over = \sin \left( {{\phi _R}\left( {\bf{q}} \right)} \right)\cos \left( {{\eta _R}\left( {\bf{q}} \right)} \right) = \frac{{{q_x}}}{{\left\| {{\bf{\bar q}}} \right\|}}$ can be interpreted as the {\it spatial frequency} along the $x$-dimension corresponding to AoAs ${{\phi _R}\left( {\bf{q}} \right)}$ and ${{\eta _R}\left( {\bf{q}} \right)}$, and ${{\bar \Omega }_R}\left( {\bf{q}} \right) \buildrel \Delta \over = \sin \left( {{\phi _R}\left( {\bf{q}} \right)} \right)\sin \left( {{\eta _R}\left( {\bf{q}} \right)} \right) = \frac{{{q_y}}}{{\left\| {{\bf{\bar q}}} \right\|}}$ as the spatial frequency along the $y$-dimension. Similarly, the transmit array response with respect to AoDs from the source node to the AIRS can be obtained, which is compactly denoted as ${{\bf{a}}_{T,s}}\left( {\bf{q}} \right)$ for convenience, with ${\left\| {{{\bf{a}}_{T,s}}\left( {\bf{q}} \right)} \right\|^2} = M$. Thus, the channel matrix from the source node to the AIRS, denoted as ${\bf{G}}\left( {\bf{q}} \right) \in {{\mathbb{C}}^{N \times M}}$, can be expressed as
 \begin{equation}
 {\bf{G}}\left( {\bf{q}} \right) = \sqrt {{\beta _{\bf{G}}}\left( {\bf{q}} \right)} {e^{ - j\frac{{2\pi {d_{\bf{G}}}}}{\lambda }}}{{\bf{a}}_R}\left( {{\phi _R}\left( {\bf{q}} \right),{\eta _R}\left( {\bf{q}} \right)} \right){\bf{a}}_{T,s}^H\left( {\bf{q}} \right),
 \end{equation}
Note that the AIRS placement $\bf{q}$ affects not only the channel power gain ${\beta _{\bf{G}}}\left( {\bf{q}} \right)$, but also the AoDs and AoAs of the source-AIRS link.

Similarly, denote by ${{\phi _T}\left( {{\bf{q}},{\bf{w}}} \right)}$ and ${{\eta _T}\left( {{\bf{q}},{\bf{w}}} \right)}$ the zenith and azimuth AoDs for the communication link from the AIRS to any location $\mathbf w\in \mathcal A$. The reflect array response at the AIRS can be similarly obtained as
\begin{equation}
\begin{aligned}
&{{\bf{a}}_T}\left( {{\phi _T}\left( {{\bf{q}},{\bf{w}}} \right),{\eta _T}\left( {{\bf{q}},{\bf{w}}} \right)} \right) = \\
&{\left[ {1, \cdots ,{e^{ - j2\pi \left( {{N_x} - 1} \right){{\bar d}_x}{{\bar \Phi} _T}\left( {{\bf{q}},{\bf{w}}} \right)}}} \right]^T} \otimes \\
&{\left[ {1, \cdots ,{e^{ - j2\pi \left( {{N_y} - 1} \right){{\bar d}_y}{{\bar \Omega} _T}\left( {{\bf{q}},{\bf{w}}} \right)}}} \right]^T}, \label{UPATransmitArrayResponseAtAIRS}
\end{aligned}
\vspace{-0.3cm}
\end{equation}
where ${{\bar \Phi }_T}\left( {{\bf{q}},{\bf{w}}} \right) \buildrel \Delta \over = \sin \left( {{\phi _T}\left( {{\bf{q}},{\bf{w}}} \right)} \right)\cos \left( {{\eta _T}\left( {{\bf{q}},{\bf{w}}} \right)} \right) = \frac{{{w_x} - {q_x}}}{{\left\| {{\bf{\bar w}} - {\bf{\bar q}}} \right\|}}$ with ${\bf{\bar w}} = {\left[ {{w_x},{w_y},0} \right]^T}$ can be interpreted as the spatial frequency along the $x$-dimension corresponding to AoDs ${{\phi _T}\left( {{\bf{q}},{\bf{w}}} \right)}$ and ${{\eta _T}\left( {{\bf{q}},{\bf{w}}} \right)}$, and ${{\bar \Omega }_T}\left( {{\bf{q}},{\bf{w}}} \right) \buildrel \Delta \over = \sin \left( {{\phi _T}\left( {{\bf{q}},{\bf{w}}} \right)} \right)\sin \left( {{\eta _T}\left( {{\bf{q}},{\bf{w}}} \right)} \right) = \frac{{{w_y} - {q_y}}}{{\left\| {{\bf{\bar w}} - {\bf{\bar q}}} \right\|}}$ as the spatial frequency along the $y$-dimension. Then the channel from the AIRS to a location $\mathbf w\in \mathcal A$, denoted as ${{\bf{h}}^H}\left( {{\bf{q}},{\bf{w}}} \right) \in {{\mathbb{C}}^{1 \times N}}$, can be expressed as
 \begin{equation}
 {{\bf{h}}^H}\left( {{\bf{q}},{\bf{w}}} \right) = \sqrt {{\beta _{\bf{h}}}\left( {{\bf{q}},{\bf{w}}} \right)} {e^{ - j\frac{{2\pi {d_{\bf{h}}}}}{\lambda }}}{\bf{a}}_T^H\left( {{\phi _T}\left( {{\bf{q}},{\bf{w}}} \right),{\eta _T}\left( {{\bf{q}},{\bf{w}}} \right)} \right).
 \end{equation}
As a result, the received signal at each location $\mathbf w\in \mathcal A$ is
  \begin{equation}
y\left( {{\bf{q}},{\bf{\Theta }},{\bf{w}},{\bf{v}}} \right) = {{\bf{h}}^H}\left( {{\bf{q}},{\bf{w}}} \right){\bf{\Theta G}}\left( {\bf{q}} \right){\bf{v}}\sqrt P s + n,
  \end{equation}
  where ${\bf{\Theta }} = {\rm{diag}}\left( {{e^{j{\theta _1}}}, \cdots ,{e^{j{\theta _N}}}} \right)$ is a diagonal phase-shift matrix with ${\theta _n} = {\theta _{{n_x},{n_y}}} = {\theta _{\left( {{n_x} - 1} \right){N_y} + {n_y}}} \in \left[ {0,2\pi } \right)$ denoting the phase shift of the $n$th reflecting  element that is located at the $n_x$th column and $n_y$th row on the IRS; $P$ and $s$ are the transmit power and information-bearing signal at the source node, respectively; ${\bf{v}} \in {{\mathbb{C}}^{M \times {\rm{1}}}}$ is the transmit beamforming vector at the source node with $\left\| {\bf{v}} \right\| = 1$; $n$ is the additive white Gaussian noise (AWGN) with zero mean and power $\sigma^2$. The received SNR at the location $\mathbf w\in \mathcal A$ is thus expressed as
  \begin{equation}
  \gamma \left( {{\bf{q}},{\bf{\Theta }},{\bf{w}},{\bf{v}}} \right) = \bar P{\left| {{{\bf{h}}^H}\left( {{\bf{q}},{\bf{w}}} \right){\bf{\Theta G}}\left( {\bf{q}} \right){\bf{v}}} \right|^2}, \label{SNRAtLocationw}
 \end{equation}
  where $\bar P = \frac{P}{\sigma ^2}$. By denoting ${\bm{\theta }} = \left[ {{\theta _1}, \cdots ,{\theta _N}} \right]$, our objective is to maximize the worst-case/minimum SNR within the target area $\mathcal A$, by jointly optimizing the transmit beamforming vector $\bf{v}$ of the source node, as well as the AIRS placement $\bf{q}$ and its 3D passive beamforming with phase shifts $\bm{\theta }$. The problem is formulated as
\begin{equation}\label{original problem}
 \begin{aligned}
\left( \rm{{P1}} \right){\rm{    }}\mathop {\max }\limits_{{\bf{q}},{\bm{\theta }},{\bf{v}}} &\  \  \mathop {\min \ \ }\limits_{{\bf{w}} \in {\rm{{\cal A}}}} \gamma \left( {{\bf{q}},{\bf{\Theta }},{\bf{w}},{\bf{v}}} \right)\\
{\rm{    s.t.       }}{\rm{  }}&\ \  {\rm{                }}0 \le {\theta _n} < 2\pi ,\ {\rm{    }}n = 1, \cdots ,N, \\
&\ \left\| {\bf{v}} \right\| = 1. \nonumber
\end{aligned}
\end{equation}

Note that in practice, an embedded micro-controller in the AIRS could communicate with the source node through a separate reliable wireless control link, which enables the instantaneous control of the AIRS~\cite{wu2020intelligentTutorial}. Furthermore, it can be seen from (P1) that with the LoS-dominating channels, the beamforming design for area coverage enhancement mainly depends on the size and location of the target area, instead of the instantaneous CSI of any particular user inside the target area~\cite{zhou2020framework,zhou2020user,wei2020channel}. With the coverage of the target area enhanced by such a design, for users located in the area, the conventional multiple access techniques such as TDMA, FDMA, and  orthogonal frequency-division multiple-access (OFDMA) can be used to separate the user transmissions. This eases the control, synchronization and channel estimation requirements compared to most existing works on IRS-aided communication based on instantaneous CSI. By exploiting the special structure of the concatenated channel ${{{\bf{\tilde h}}}^H} \triangleq {{\bf{h}}^H}\left( {{\bf{q}},{\bf{w}}} \right){\bf{\Theta G}}\left( {\bf{q}} \right)$, we first show that the optimal transmit beamforming vector $\mathbf v$ corresponds to the simple MRT towards the AIRS, regardless of the reflecting link from the AIRS to the target area.
\begin{proposition}
 The optimal transmit beamforming vector $\bf{v}$ to (P1) is ${{\bf{v}}^*} = \frac{{{{\bf{a}}_{T,s}}\left( {\bf{q}} \right)}}{{\sqrt M }}$.
\end{proposition}
\begin{IEEEproof}
For any given AIRS placement $\mathbf q$, target location $\mathbf w$ and AIRS phase shifts $\boldsymbol \theta$, it is known that the optimal transmit beamforming vector to maximize $\gamma \left( {{\bf{q}},{\bf{\Theta }},{\bf{w}},{\bf{v}}} \right)$ in (\ref{SNRAtLocationw}), denoted as ${{\bf{v}}^*}\left( {{\bf{q}},{\bf{\Theta }},{\bf{w}}} \right)$, is the eigenvector corresponding to the largest eigenvalue of the channel matrix ${\bf{\tilde h}}{{{\bf{\tilde h}}}^H}$. Furthermore, ${\bf{\tilde h}}{{{\bf{\tilde h}}}^H}$ can be simplified as
  \begin{equation}
  \small
  \begin{aligned}
  &{\bf{\tilde h}}{{{\bf{\tilde h}}}^H} = {{\bf{G}}^H}\left( {\bf{q}} \right){{\bf{\Theta }}^H}{\bf{h}}\left( {{\bf{q}},{\bf{w}}} \right){{\bf{h}}^H}\left( {{\bf{q}},{\bf{w}}} \right){\bf{\Theta G}}\left( {\bf{q}} \right)\\
  &= {\beta _{\bf{G}}}\left( {\bf{q}} \right){\left| {{\bf{a}}_R^H\left( {{\phi _R}\left( {\bf{q}} \right),{\eta _R}\left( {\bf{q}} \right)} \right){{\bf{\Theta }}^H}{\bf{h}}\left( {{\bf{q}},{\bf{w}}} \right)} \right|^2}{{\bf{a}}_{T,s}}\left( {\bf{q}} \right){\bf{a}}_{T,s}^H\left( {\bf{q}} \right). \label{hhExpression}
  \end{aligned}
 \end{equation}
It readily follows that ${\bf{\tilde h}}{{{\bf{\tilde h}}}^H}$ is a rank-one matrix, whose eigenvector is  ${{\bf{v}}^*} = \frac{{{{\bf{a}}_{T,s}}\left( {\bf{q}} \right)}}{{\sqrt M }}$. More importantly, this eigenvector is independent of the target location $\mathbf w$. Thus, regardless of $\mathbf w$, it is optimal to set the transmit beamforming as  ${{\bf{v}}^*} = \frac{{{{\bf{a}}_{T,s}}\left( {\bf{q}} \right)}}{{\sqrt M }}$ in (P1). The proof is thus completed.
\end{IEEEproof}

By substituting the optimal ${\bf{v}}^*$ into (\ref{SNRAtLocationw}) and after some manipulations, the corresponding SNR at the target location $\mathbf w\in \mathcal A$ can be written as \eqref{SNR at location w}, shown at the top of the next page.  As a result, problem (P1) reduces to
\begin{equation}
\begin{aligned}
 \left( {\rm{P2}} \right)\ {\rm{    }}\mathop {\max }\limits_{{\bf{q}},{\bm{\theta }}}&\  \   \mathop {\min }\limits_{{\bf{w}} \in {\rm{{\cal A}}}} \ \gamma \left( {{\bf{q}},{\bf{\Theta }},{\bf{w}}} \right)\\
{\rm{      s.t.     }}{\rm{  }}&\ \  0 \le {\theta _n} < 2\pi ,\ {\rm{    }}n = 1, \cdots ,N. \label{problem without v} \nonumber
\end{aligned}
\end{equation}
Problem (P2) is difficult to be directly solved due to the following reasons. First, the objective function is the worst-case SNR over a two-dimensional (2D) area, which is difficult to be explicitly expressed in terms of the optimization variables. Second, the optimization problem is highly non-convex and the optimization variables $\bf{q}$ and $\bm{\theta}$ are intricately coupled with each other, as shown in \eqref{SNR at location w}. To tackle this problem, we first consider the special case of (P2) for a given location ${\bf{w}} \in {\cal A}$, for which the optimal solutions for the AIRS phase shifts and placement are derived in closed-form. Then for the general case of (P2), we first consider the simplified ULA-based AIRS, i.e., $N_y=1$ and $N ={N_x}$, where the passive beamforming only involves the beam steering for the spatial frequency $\bar \Phi $ along the $x$-dimension. However, even for this simplified case, problem (P2) is still non-convex and difficult to solve. We thus propose an efficient suboptimal solution by decoupling the 2D passive beamforming design of the AIRS phase-shift vector ${\bm {\theta}}$ and its placement optimization $\bf{q}$, with the former aiming to maximize the worst-case array gain and the latter to balance the angular span and the cascaded path loss. In particular, sub-array based beam flattening technique is applied to form one flattened beam with its beamwidth catering to the size of the target area, thus achieving an approximately equal array gain for all locations in it. Finally, the proposed design for ULA-based AIRS is extended to the general case of UPA-based AIRS, which enables 3D passive beam flattening to cover the target area efficiently.
\newcounter{mytempeqncnt1}
\begin{figure*}
\normalsize
\setcounter{mytempeqncnt1}{\value{equation}}
\begin{align}
 \gamma \left( {{\bf{q}},{\bf{\Theta }},{\bf{w}}} \right) &= \bar P{\beta _{\bf{G}}}\left( {\bf{q}} \right)M{\left| {{{\bf{h}}^H}\left( {{\bf{q}},{\bf{w}}} \right){\bf{\Theta }}{{\bf{a}}_R}\left( {{\phi _R}\left( {\bf{q}} \right),{\eta _R}\left( {\bf{q}} \right)} \right)} \right|^2}\notag\\
 & = \frac{{\bar P\beta _0^2M{{\left| {\sum\limits_{{n_x} = 1}^{{N_x}} {\sum\limits_{{n_y} = 1}^{{N_y}} {{e^{j2\pi \left( {{n_x} - 1} \right){{\bar d}_x}\left[ {{{\bar \Phi }_T}\left( {{\bf{q}},{\bf{w}}} \right) - {{\bar \Phi }_R}\left( {\bf{q}} \right)} \right]}}{e^{j2\pi \left( {{n_y} - 1} \right){{\bar d}_y}\left[ {{{\bar \Omega }_T}\left( {{\bf{q}},{\bf{w}}} \right) - {{\bar \Omega }_R}\left( {\bf{q}} \right)} \right]}}{e^{j{\theta _{{n_x},{n_y}}}}}} } } \right|}^2}}}{{\left( {{H^2} + {{\left\| {{\bf{q}} - {\bf{w}}} \right\|}^2}} \right)\left( {{H^2} + {{\left\| {\bf{q}} \right\|}^2}} \right)}}. \label{SNR at location w}
\end{align}
\hrulefill 
\vspace*{4pt} 
\end{figure*}
\section{Optimization for SNR Maximization at Single Target Location}\label{SingleLocationSNREnhancementCase}
In this section, we consider the special case of (P2) where $\mathcal A$ degenerates to one single point, denoted as ${{\bf{w}}_1}$. In this case, the inner minimization of the objective function in (P2) is irrelevant, and problem (P2) reduces to
\begin{equation}
\begin{aligned}
\left( {{\rm{P}}3}\right){\rm{    }}\mathop {\max }\limits_{{{\bf{q}}},{\bm{\theta }}} &\  \  {\gamma_1} \left( {{\bf{q}},{\bf{\Theta }}} \right)\\
{\rm{           }}{\rm{s.t.}}{\rm{  }}&\ \ 0 \le {\theta _{n}} < 2\pi ,\ {\rm{    }}n = 1, \cdots ,N. \label{single location problem} \nonumber
\end{aligned}
\end{equation}
 It is not difficult to see that at the optimal solution to (P3), the different rays reflected by the AIRS should be coherently added at the designated location ${{\bf{w}}_1}$. Therefore, based on \eqref{SNR at location w}, for any given AIRS placement ${\bf{q}}$, the optimal phase shifts for passive beamforming are given by
\begin{equation}
\begin{aligned}
\theta _{{n_x},{n_y}}^*\left( {\bf{q}} \right)&= \bar \theta  - 2\pi \left( {{n_x} - 1} \right){{\bar d}_x}\left[ {{{\bar \Phi }_T}\left( {{\bf{q}},{{\bf{w}}_1}} \right) - {{\bar \Phi }_R}\left( {\bf{q}} \right)} \right]\\
&- 2\pi \left( {{n_y} - 1} \right){{\bar d}_y}\left[ {{{\bar \Omega }_T}\left( {{\bf{q}},{{\bf{w}}_1}} \right) - {{\bar \Omega }_R}\left( {\bf{q}} \right)} \right],  \label{single location optimal phase shift}
\end{aligned}
\end{equation}
where $1 \le {n_x} \le {N_x}$, $1 \le {n_y} \le {N_y}$, and $\bar \theta$ is an arbitrary phase shift that is common to all reflecting elements. As a result, the received SNR at the target location ${\bf w}_1$ is simplified as
\begin{equation}
{\gamma _1}\left( {\bf{q}} \right) = \frac{{\bar P\beta _0^2M{N^2}}}{{\left( {{H^2} + {{\left\| {{{\bf{q}}} - {{\bf{w}}_1}} \right\|}^2}} \right)\left( {{H^2} + {{\left\| {{{\bf{q}}}} \right\|}^2}} \right)}}.\label{single-location SNR}
 \end{equation}
To maximize the received SNR given in \eqref{single-location SNR} at the single target location ${\bf{w}}_1$, problem (P3) reduces to
\begin{equation}
\vspace{-0.2cm}
\left( {\rm{P4}} \right){\rm{    }}\mathop {\min }\limits_{{{\bf{q}}}} {\rm{  }}\left( {{H^2} + {{\left\| {{{\bf{q}}} - {{\bf{w}}_1}} \right\|}^2}} \right)\left( {{H^2} + {{\left\| {{{\bf{q}}}} \right\|}^2}} \right).\nonumber
 \end{equation}
\begin{proposition}
For the single-location SNR maximization problem (P4), the optimal AIRS 2D placement solution is given by
 \begin{equation}
  {\bf{q}}^* =   {\xi ^*}\left( \rho  \right){{\bf{w}}_1} , \label{singleLocationOptimalPlacement}
  \end{equation}
 where $\rho  \buildrel \Delta \over = \frac{{\left\| {{{\bf{w}}_1}} \right\|}}{H}$, and
 \begin{equation}
 {\xi ^*}\left( \rho  \right) = \left\{ \begin{aligned}
 &\frac{1}{2},\ \ \ \ \ \ \ \ \ \ \ \ \ \ \ \ \ \ \ \ {\rm{                      if  }}\ \ 0 \le \rho  \le 2\\
 &\frac{1}{2} \pm \sqrt {\frac{1}{4} - \frac{1}{{{\rho ^2}}}},\ \ \ \ {\rm{     otherwise}}.
 \end{aligned} \right.\label{optimal placement for single-location}
 \end{equation}
 \label{PropositionoptimalPlacementsinglelocation}
\end{proposition}
 \begin{IEEEproof}
 The result can be obtained by checking the first-order derivative, which is omitted for brevity.
 \end{IEEEproof}

\begin{figure}[!h]
  \centering
  \centerline{\includegraphics[width=3.5in,height=2.625in]{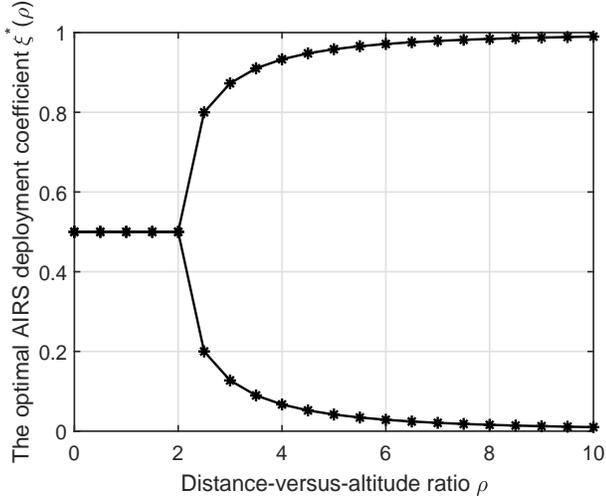}}
  \caption{The optimal AIRS deployment coefficient $  {\xi ^*}\left( \rho  \right)$ against distance-versus-altitude ratio $\rho {\rm{ = }}\frac{{\left\| {{{\bf{w}}_1}} \right\|}}{H}$.}
  \label{coefficientVersusRatioRho}
  \end{figure}

  Proposition~\ref{PropositionoptimalPlacementsinglelocation} shows that the optimal horizontal placement of the AIRS only depends on $\rho$, i.e., the ratio between the source-destination distance ${\left\| {{{\bf{w}}_1}} \right\|}$ and AIRS altitude $H$.
  For $0 \le \rho  \le 2$, the AIRS should always be placed exactly above the midpoint between the source node and the target location. On the other hand, for $\rho  > 2$, there exist two optimal horizontal AIRS placement locations that are symmetric over the midpoint, as shown in Fig.~\ref{coefficientVersusRatioRho}. Note that the above result is different from the conventional active relay placement \cite{han2013relay}, whose optimal solution generally depends on the transmit power and the relay processing noise power.

  With the optimal AIRS placement location in \eqref{singleLocationOptimalPlacement}, the optimal SNR at the single target location ${\bf w}_1$ is given by \eqref{singleLocationOptimalSNR}, shown at the top of the next page.
  \newcounter{mytempeqncnt2}
\begin{figure*}
\normalsize
\setcounter{mytempeqncnt2}{\value{equation}}
\begin{equation}
  \gamma _1^*\left( \rho  \right) = \left\{ \begin{aligned}
  &\frac{{\bar P\beta _0^2M{N^2}}}{{{{\left( {{H^2} + \frac{1}{4}{{\left\| {{{\bf{w}}_1}} \right\|}^2}} \right)}^2}}},\ \ \ \ \ \ \ \ \ \ \ \ \ \ \ \ \ \ \ \ \ \ \ \ \ \ \ \ \ \ \ \ \ \ \ \ \ \ \ \ \ \ \ \ \ \ \ \ \ \ \ \ \ \ \ \ \ \ \ \ \ \ {\rm{                      if  }} \ \ 0 \le \rho  \le {\rm{2}}\\
  &\frac{{\bar P\beta _0^2M{N^2}}}{{\left( {{H^2} + {{\left( {\frac{1}{2} + \sqrt {\frac{1}{4} - \frac{1}{{{\rho ^2}}}} } \right)}^2}{{\left\| {{{\bf{w}}_1}} \right\|}^2}} \right)\left( {{H^2} + {{\left( {\frac{1}{2} - \sqrt {\frac{1}{4} - \frac{1}{{{\rho ^2}}}} } \right)}^2}{{\left\| {{{\bf{w}}_1}} \right\|}^2}} \right)}},\ \ {\rm{     otherwise}}. \label{singleLocationOptimalSNR}
  \end{aligned} \right.
\end{equation}
\hrulefill 
\vspace*{4pt} 
\end{figure*}
\begin{remark}
It is observed from \eqref{singleLocationOptimalSNR} that since the optimal SNR decreases with the increase of the AIRS altitude $H$, a relatively small $H$ should be selected to enhance the performance. However, such a result is based on the LoS link assumptions. In order to obtain LoS channels with both source/target location, we need to choose sufficiently large $H$. Therefore, this is also an important trade-off for AIRS deployment in practice for altitude selection.
\end{remark}
\section{Optimization for Min-SNR Maximization in Area Coverage}
In this section, we study the general case of (P2) for AIRS-enabled area coverage. To gain useful insights, we first consider the special case of ULA-based AIRS, i.e., $N_y=1$ and $N ={N_x}$, for which passive beamforming is only applied to the $x$-dimension for beam steering over the spatial frequency $\bar \Phi $. Then, we extend our proposed solution to the general case of UPA-based AIRS with 3D passive beamforming over both the spatial frequency dimensions of $\bar \Phi$ and $\bar \Omega$.
\subsection{The Special Case of ULA-Based AIRS}
For ULA-based AIRS, by substituting $N_y=1$ and $N ={N_x}$ into \eqref{SNR at location w}, the SNR reduces to
\begin{equation}
\gamma \left( {{\bf{q}},{\bf{\Theta }},{\bf{w}}} \right) = \frac{{\bar P\beta _0^2M{{\left| {\sum\limits_{n = 1}^N {{e^{j\left( {{\theta _n} + 2\pi \left( {n - 1} \right){{\bar d}_x}\left[ {{{\bar \Phi }_T}\left( {{\bf{q}},{\bf{w}}} \right) - {{\bar \Phi }_R}\left( {\bf{q}} \right)} \right]} \right)}}} } \right|}^2}}}{{\left( {{H^2} + {{\left\| {{\bf{q}} - {\bf{w}}} \right\|}^2}} \right)\left( {{H^2} + {{\left\| {\bf{q}} \right\|}^2}} \right)}}, \label{ULASNRAtW}
\end{equation}
where ${\theta _n} = {\theta _{{n_x}}},1 \le n \le N$.

Note that even for the ULA case with the SNR given by \eqref{ULASNRAtW}, solving problem (P2) by standard optimization techniques is difficult in general. By exploiting the fact that the phase shifts optimization for IRS resembles the extensively studied phase array design or analog beamforming, we propose an efficient two-step solution to (P2) by decoupling the phase shifts optimization and AIRS placement optimization. First, it is noted that by discarding constant terms, problem (P2) with $\gamma \left( {{\bf{q}},{\bf{\Theta }},{\bf{w}}} \right)$ given in \eqref{ULASNRAtW} can be equivalently written as
\begin{equation}
\begin{aligned}
 \left( {\rm{P5}} \right)\ {\rm{    }}\mathop {\max }\limits_{{\bf{q}},{\bm{\theta }}}&\  \   \mathop {\min }\limits_{{\bf{w}} \in {\rm{{\cal A}}}} \ \frac{{{f_1}\left( {{\bf{q}},{\bm{\theta }},{\bf{w}}} \right)}}{{{f_2}\left( {{\bf{q}},{\bf{w}}} \right)}}\\
{\rm{      s.t.     }}{\rm{  }}&\ \  0 \le {\theta _n} < 2\pi ,\ {\rm{    }}n = 1, \cdots ,N, \label{problem without v}\nonumber
\end{aligned}
\end{equation}
where ${f_1}\left( {{\bf{q}},{\bm{\theta }},{\bf{w}}} \right) \buildrel \Delta \over = {\left| {\sum\limits_{n = 1}^N {{e^{j\left( {{\theta _n} + 2\pi \left( {n - 1} \right){{\bar d}_x}\left[ {{{\bar \Phi }_T}\left( {{\bf{q}},{\bf{w}}} \right) - {{\bar \Phi }_R}\left( {\bf{q}} \right)} \right]} \right)}}} } \right|^2}$ accounts for the {\it array gain} due to the passive beamforming by the AIRS,  and ${f_2}\left( {{\bf{q}},{\bf{w}}} \right) \buildrel \Delta \over = \left( {{H^2} + {{\left\| {{\bf{q}} - {\bf{w}}} \right\|}^2}} \right)\left( {{H^2} + {{\left\| {\bf{q}} \right\|}^2}} \right)$ accounts for the concatenated path loss from the source node to a location ${\bf{w}} \in {\cal A}$ via the AIRS located at $\bf q$.

With the proposed solution, for any given AIRS placement $\mathbf q$, the phase shifts in $\boldsymbol \theta$ are designed to maximize the minimum/worst-case array gain among all locations in $\cal A$ by solving the following problem
\begin{equation}
\begin{aligned}
 \left({\rm{P5.1}}\right)\   \mathop {\max }\limits_{\bm{\theta }} &\ \ \mathop {\min }\limits_{{\bf{w}} \in {\rm{{\cal A}}}}\  {f_1}\left( {{\bf{q}},{\bm{\theta }},{\bf{w}}} \right) \\
 {\rm{      s.t.     }}{\rm{  }}&\ \  0 \le {\theta _n} < 2\pi ,\ {\rm{    }}n = 1, \cdots ,N. \label{step1 problem}\nonumber
 \end{aligned}
\end{equation}
Note that (P5.1) is an approximation of the original problem (P5) with given $\mathbf q$, since ${f_2}\left( {{\bf{q}},{\bf{w}}} \right)$ is ignored in the inner minimization of the objective function in (P5.1). Such an approximation is reasonable since the array gain ${f_1}\left( {{\bf{q}},\bm{\theta},{\bf{w}}} \right)$ is usually more sensitive than the concatenated path loss ${f_2}\left( {{\bf{q}},{\bf{w}}} \right)$ to the location variation of $\mathbf w$ in the target area $\mathcal A$, especially when $\mathcal A$ is small or $H \gg {D_x}$ and ${D_y}$. After solving (P5.1), in the second step, the obtained solution to (P5.1), denoted as $ {{\bm{\theta }}^*}\left( {\bf{q}} \right)$, is substituted into the objective function of (P5), based on which the AIRS placement is optimized by solving the following problem
\begin{equation}
 \left({\rm{P5.2}}\right)\ \mathop {\max }\limits_{\bf{q}} \ \mathop {\min }\limits_{{\bf{w}} \in {\rm{{\cal A}}}} \ \frac{{{f_1}\left( {{\bf{q}},{{\bm{\theta }}^*}\left( {\bf{q}} \right),{\bf{w}}} \right)}}{{{f_2}\left( {{\bf{q}},{\bf{w}}} \right)}}.\nonumber
 \end{equation}
As such, the obtained max-min SNR within the target area $\mathcal A$ after solving the two-step problems will be a lower bound of the optimal value of (P5). In the following, we present the details for solving (P5.1) and (P5.2), respectively.
\subsubsection{Beam Broadening and Flattening for Passive Beamforming}
For any given AIRS placement $\mathbf q$, let ${\Delta _{\min }}\left( {\bf{q}} \right)$ and ${\Delta _{\max }}\left( {\bf{q}} \right)$ denote the minimum and maximum deviation of the spatial frequency ${{{\bar \Phi }_T}\left( {{\bf{q}},{\bf{w}}} \right)}$ corresponding to the AoDs along the $x$-axis from ${{\bar \Phi} _R}\left( {\bf{q}} \right)$ in the target area $\mathcal A$, respectively, i.e.,
\begin{equation}
{\Delta _{\min }}\left( {\bf{q}} \right) \buildrel \Delta \over = \mathop {\min }\limits_{{\bf{w}} \in A} {{\bar \Phi }_T}\left( {{\bf{q}},{\bf{w}}} \right) - {{\bar \Phi }_R}\left( {\bf{q}} \right), \label{theMinimumDeviation}
\end{equation}
\begin{equation}
{\Delta _{\max }}\left( {\bf{q}} \right) \buildrel \Delta \over = \mathop {\max }\limits_{{\bf{w}} \in {\rm{{\cal A}}}} {\rm{ }}{ {\bar \Phi} _T}\left( {{\bf{q}},{\bf{w}}} \right) - {{\bar \Phi} _R}\left( {\bf{q}} \right). \label{theMaximumDeviation}
\end{equation}
Then (P5.1) can be equivalently written as
\begin{equation}
\small
\begin{aligned}
\mathop {\max}\limits_{\bm{\theta }} &\ \mathop {\min }\limits_{{\Delta _{\min }}\left( {\bf{q}} \right) \le \Delta  \le {\Delta _{\max }}\left( {\bf{q}} \right)} \ {g_1}\left( {{\bm{\theta }},\Delta } \right) \buildrel \Delta \over = {\left| {\sum\limits_{n = 1}^N {{e^{j\left( {{\theta _n} + 2\pi \left( {n - 1} \right){\bar d}_x \Delta } \right)}}} } \right|^2} \\
{\rm{                  s}}{\rm{.t}}{\rm{.      }}&\ \ 0 \le {\theta _n} < 2\pi,\ 1 \le n \le N.\label{phaseShiftsOptimizationwithArrayGaing1}
 \end{aligned}
\end{equation}
While directly solving the above optimization problem is challenging due to its non-convexity, we propose an efficient solution based on the beam flattening technique. Specifically, for the above max-min problem, $\bm{\theta}$ should be designed such that the array gain ${g_1}\left( {{\bm{\theta }},\Delta } \right)$ defined in \eqref{phaseShiftsOptimizationwithArrayGaing1} is approximately equal for all $\Delta$ within the interval $\left[ {{\Delta _{\min }}\left( {\bf{q}} \right),{\Delta _{\max }}\left( {\bf{q}} \right)} \right]$, which implies a flattened beam pattern of the AIRS in this interval. Towards this end, the $N$-element array is partitioned into $L$ sub-arrays with ${N_s} = N/L$ elements in each sub-array. For notational convenience, we assume that $N/L$ is an integer. By using the array manifold concept, the design of phase shifts vector $\bm{\theta}$ can be transformed into that of AIRS's array manifold. Specifically, the manifold of the $l$th sub-array, denoted as ${{\bf{a}}_l}\left( {{{\bar \Phi }_l}} \right)$, aims to direct its sub-beam towards the spatial frequency direction ${{{\bar \Phi }_l}}$, which can be expressed as
\begin{equation}
\small{
{{\bf{a}}_l}\left( {{{\bar \Phi }_l}} \right) = {e^{j{\alpha _l}}}\left[ {1,{e^{ - j2\pi {{\bar d}_x}{{\bar \Phi }_l}}}, \cdots ,} \right.{\left. {{e^{ - j2\pi \left( {{N_s} - 1} \right){{\bar d}_x}{{\bar \Phi }_l}}}} \right]^T},l = 1, \cdots ,L,} \label{subArrayManifold}
\end{equation}
where ${{\bar \Phi} _l}{\rm{ }} \buildrel \Delta \over = \sin \left( {{\phi _l}} \right)\cos \left( {{\eta _l}} \right)$ with ${\phi _l}$ and ${\eta _l}$ denoting the zenith and azimuth angles of beam direction for the $l$th sub-array, respectively; ${e^{j{\alpha _l}}}$ is a common phase term for the $l$th sub-array. Thus, the phase shifts ${\theta _n}$ in \eqref{phaseShiftsOptimizationwithArrayGaing1} can be expressed as ${\theta _n} = {\theta _{\left( {l - 1} \right){N_s} + i}} = {\alpha _l} - 2\pi \left( {i - 1} \right){\bar d}_x{{\bar \Phi }_l},l = 1, \cdots ,L,i = 1, \cdots ,{N_s}$. Then the array gain ${g_1}\left( {{\bm{\theta }},\Delta } \right)$ in \eqref{phaseShiftsOptimizationwithArrayGaing1} can be further expressed as a function of $L$, $N_s$ and $\left\{ {{\alpha _l},{{\bar \Phi }_l}} \right\}_{l = 1}^L$, i.e.,
\begin{equation}
\hspace{-0.3cm}
\small
\begin{aligned}
&{g_2}\left( {L,{N_s},\left\{ {{\alpha _l},{{\bar \Phi }_l}} \right\},\Delta } \right)\\
 &= {\left| {\sum\limits_{l = 1}^L {\sum\limits_{i = 1}^{{N_s}} {{e^{j{\theta _{\left( {l - 1} \right){N_s} + i}}}}{e^{j2\pi \left( {\left( {l - 1} \right){N_s} + i - 1} \right){{\bar d}_x}\Delta }}} } } \right|^2}\\
 &= {\left| {\sum\limits_{l = 1}^L {{e^{j\left( {{\alpha _l} + 2\pi \left( {l - 1} \right){N_s}{{\bar d}_x}\Delta } \right)}}\sum\limits_{i = 1}^{{N_s}} {{e^{j2\pi \left( {i - 1} \right){{\bar d}_x}\left( {\Delta  - {{\bar \Phi }_l}} \right)}}} } } \right|^2}\\
 &= {\left| {\sum\limits_{l = 1}^L {{e^{j\left( {{\alpha _l} + \pi \left( {\left( {2l - 1} \right){N_s} - 1} \right){{\bar d}_x}\Delta  - \pi \left( {{N_s} - 1} \right){{\bar d}_x}{{\bar \Phi }_l}} \right)}}\frac{{\sin \left( {\pi {N_s}{{\bar d}_x}\left( {\Delta  - {{\bar \Phi }_l}} \right)} \right)}}{{\sin \left( {\pi {{\bar d}_x}\left( {\Delta  - {{\bar \Phi }_l}} \right)} \right)}}} } \right|^2}\\
 &= {\left| {\sum\limits_{l = 1}^L {{e^{jh\left( {{\alpha _l},\Delta ,{{\bar \Phi }_l}} \right)}}\frac{{\sin \left( {\pi {N_s}{{\bar d}_x}\left( {\Delta  - {{\bar \Phi }_l}} \right)} \right)}}{{\sin \left( {\pi {{\bar d}_x}\left( {\Delta  - {{\bar \Phi }_l}} \right)} \right)}}} } \right|^2},\label{arrayGaing2}
\end{aligned}
\end{equation}
where $h\left( {{\alpha _l},\Delta ,{{\bar \Phi }_l}} \right) \buildrel \Delta \over = {\alpha _l} + \pi \left( {\left( {2l - 1} \right){N_s} - 1} \right){{\bar d}_x}\Delta  - \pi \left( {{N_s} - 1} \right){{\bar d}_x}{{ \bar \Phi }_l}$. As such, problem~\eqref{phaseShiftsOptimizationwithArrayGaing1} can be transformed to
\begin{equation}
\begin{aligned}
\mathop {\max }\limits_{L,{N_s},\left\{ {{\alpha _l},{{\bar \Phi }_l}} \right\}} &\ \mathop {\min }\limits_{{\Delta _{\min }}\left( {\bf{q}} \right) \le \Delta  \le {\Delta _{\max }}\left( {\bf{q}} \right)} {g_2}\left( {L,{N_s},\left\{ {{\alpha _l},{{\bar \Phi }_l}} \right\},\Delta } \right)\\
{\rm{                  s}}{\rm{.t}}{\rm{.      }}&\ \ L{N_s} = N,\\
&\ \  - 1 \le {{\bar \Phi }_l} \le 1,\ l = 1, \cdots ,L,\\
&\ \ 0 \le {\alpha _l} < 2\pi,\ l = 1, \cdots ,L. \label{phaseShiftOptimizationEquivalentProblem}
\end{aligned}
\end{equation}

\begin{figure}[!t]
  \centering
  \centerline{\includegraphics[width=3.0in, height=1.05in]{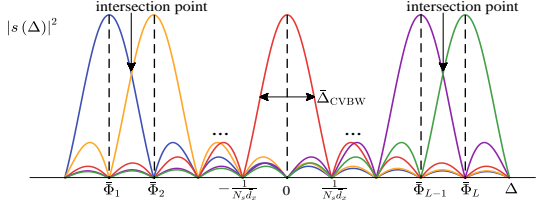}}
  \caption{An illustration of beam flattening via sub-array grouping and steering.}
  \label{subArrayBeamDirection}
  \end{figure}

 To solve \eqref{phaseShiftOptimizationEquivalentProblem}, we first study the property of the following function that appears in \eqref{arrayGaing2}:
  \begin{equation}
  s\left( \Delta  \right) = \frac{{\sin \left( {\pi {N_s}{{\bar d}_x}\Delta } \right)}}{{\sin \left( {\pi {{\bar d}_x}\Delta } \right)}}. \label{array gain}
  \end{equation}
 It is observed that $s\left( \Delta  \right)$ has a peak at $\Delta  = 0$ with $s\left( 0 \right) = {N_s}$, and nulls at $\Delta = \pm \frac{k}{{{N_s}{{\bar d}_x}}}, k = 1, \cdots ,{N_s} - 1$. The null-to-null beamwidth is thus given by ${{\bar \Delta }_{{\rm{BW}}}} = \frac{2}{{{N_s}{{\bar d}_x}}}$, as illustrated in Fig.~\ref{subArrayBeamDirection}. This corroborates the fact that the beamwidth of phase array is inversely proportional to the array aperture ${{N_s}{{\bar d}_x}}$~\cite{mailloux2005phased,rajagopal2012beam,Orfanidis2016Electromagnetic}. Furthermore, we take the
 spatial frequency interval $\left[ { - \frac{1}{{2{N_s}{{\bar d}_x}}},\frac{1}{{2{N_s}{{\bar d}_x}}}} \right]$ as the beam coverage interval~\cite{xiao2016hierarchical}, so that the coverage beamwidth (CVBW) with sub-array size $N_s$ is
 \begin{equation}
 {{\bar \Delta }_{{\rm{CVBW}}}}\left( {{N_s}} \right) = \frac{1}{{{N_s}{{\bar d}_x}}}. \label{coverageBeamwidth}
 \end{equation}

 Next, it is observed that with the array manifold design given in \eqref{subArrayManifold}, the resulting array gain in \eqref{arrayGaing2} is given by a superposition of $L$ copies of $s\left( \Delta  \right)$, each shifted by a spatial frequency ${{\bar \Phi }_l}$ with a phase coefficient ${h\left( {{\alpha _l},\Delta ,{{\bar \Phi }_l}} \right)}$. To obtain a flattened beam, the sub-array beam directions ${{\bar \Phi }_l},\ l = 1, \cdots ,L$, should be carefully designed. Inspired by the subcarrier spacing for Orthogonal Frequency Division Multiplexing (OFDM), as illustrated in Fig.~\ref{subArrayBeamDirection}, ${{\bar \Phi }_l}$ is designed so that the adjacent spatial frequency shift is separated by the spatial frequency resolution of the subarray, namely $\frac{1}{{{N_s}{{\bar d}_x}}}$. Thus, we have
\begin{equation}
{{\bar \Phi} _l} = {{\bar \Phi} _1} + \frac{{l - 1}}{{{N_s}{{\bar d}_x}}},\ l = 1, \cdots ,L, \label{sub-arrayBeamDirection}
\end{equation}
 for some starting spatial frequency ${\bar \Phi}_1$, and the resulting beam pattern will be centered at 0 when ${{\bar \Phi }_1} =  - \frac{{L - 1}}{{2{N_s}{{\bar d}_x}}}$. Furthermore, it is seen that at $\Delta  = {{\bar \Phi }_l}$, only the $l$th sub-array would contribute to the array gain since it corresponds to the nulls of all other sub-arrays. By substituting \eqref{sub-arrayBeamDirection} into \eqref{arrayGaing2}, the resulting array gain for $\Delta  = {{\bar \Phi }_l}$ is given by ${g_2}\left( {L,{N_s},\left\{ {{\alpha _l},{{\bar \Phi }_l}} \right\},\Delta  = {{\bar \Phi }_l}} \right) = {\left| {\frac{{\sin \left( {\pi {N_s}{{\bar d}_x}\left( {\Delta  - {{\bar \Phi }_l}} \right)} \right)}}{{\sin \left( {\pi {{\bar d}_x}\left( {\Delta  - {{\bar \Phi }_l}} \right)} \right)}}} \right|^2} = N_s^2,l = 1, \cdots ,L$.

 For $\Delta  \ne {{\bar \Phi }_l}$, the array gain is in general given by the superposition of the contributions from all the $L$ sub-arrays. However, since $\frac{{\sin \left( {\pi {N_s}{{\bar d}_x}\Delta } \right)}}{{\sin \left( {\pi {{\bar d}_x}\Delta } \right)}}$ is relatively small for $|\Delta | \ge \frac{1}{{{N_s}{{\bar d}_x}}}$, a closer look at \eqref{arrayGaing2} and \eqref{sub-arrayBeamDirection} reveals that for any given $\Delta$ value, those two adjacent sub-arrays would have the most significant contributions to the array gain. Furthermore, in order to mitigate the array fluctuation, the phase shift ${\alpha _l}$ of each sub-array is designed so that the resulting array gain at
 the intersection points of the beam pattern of two adjacent sub-arrays is maximized. Specifically, for the intersection point ${\Delta _p} = {{\bar \Phi }_1} + \frac{{p - 1}}{{{N_s}{{\bar d}_x}}} + \frac{1}{{2{N_s}{{\bar d}_x}}},p = 1, \cdots ,L - 1$, the array gain can be expressed as \eqref{arrayGainAlongIntersectionPoints}, shown at the top of the next page, where in ${\left( a \right)}$ we have only retained the two most significant adjacent terms for $l=p, p+1$, since when $\left| {{\Delta _p} - {{\bar \Phi }_l}} \right| > \frac{1}{{{N_s}{{\bar d}_x}}}$, the effect of the $l$th sub-array on the intersection point $\Delta_p$ is small, as shown in Fig.~\ref{subArrayBeamDirection}. To maximize \eqref{arrayGainAlongIntersectionPoints}, we should have $h\left( {{\alpha _p},{\Delta _p},{{\bar \Phi }_{_p}}} \right) = h\left( {{\alpha _{p + 1}},{\Delta _p},{{\bar \Phi }_{_{p + 1}}}} \right) + 2k\pi ,k \in \mathbb{Z}$. By substituting ${{\bar \Phi }_p}$ and ${{\bar \Phi }_{p + 1}}$ into this equation, it follows that
   \begin{figure}[!t]
  \centering
  \centerline{\includegraphics[width=3.5in,height=2.625in]{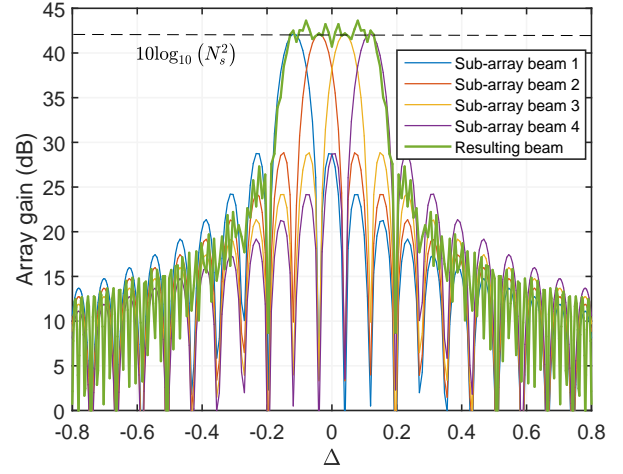}}
  \caption{Array gain of the broadened and flattened beam versus spatial frequency.}
  \label{resultingBroadenedBeam}
  \end{figure}
  \newcounter{mytempeqncnt3}
\begin{figure*}
\normalsize
\setcounter{mytempeqncnt3}{\value{equation}}
\begin{align}
 &{g_2}\left( {L,{N_s},\left\{ {{\alpha _l},{{\bar \Phi }_l}} \right\},{\Delta _p} = {{\bar \Phi }_1} + \frac{{2p - 1}}{{2{N_s}{{\bar d}_x}}}} \right){\rm{ = }}{\left| {\sum\limits_{l = 1}^L {{e^{jh\left( {{\alpha _l},{\Delta_p} ,{{\bar \Phi }_l}} \right)}}\frac{{\sin \left( {\pi {N_s}{{\bar d}_x}\left( {\frac{{2\left( {p - l} \right) + 1}}{{2{N_s}{{\bar d}_x}}}} \right)} \right)}}{{\sin \left( {\pi {{\bar d}_x}\left( {\frac{{2\left( {p - l} \right) + 1}}{{2{N_s}{{\bar d}_x}}}} \right)} \right)}}} } \right|^2}\notag\\
 &\mathop  \approx \limits^{\left( a \right)} {\left| {{e^{jh\left( {{\alpha _p},{\Delta _p},{{\bar \Phi }_{_p}}} \right)}}\frac{{\sin \left( {\pi {N_s}{{\bar d}_x}\frac{1}{{2{N_s}{{\bar d}_x}}}} \right)}}{{\sin \left( {\pi {{\bar d}_x}\frac{1}{{2{N_s}{{\bar d}_x}}}} \right)}} + {e^{jh\left( {{\alpha _{p + 1}},{\Delta _p},{{\bar \Phi }_{_{p + 1}}}} \right)}}\frac{{\sin \left( { - \pi {N_s}{{\bar d}_x}\frac{1}{{2{N_s}{{\bar d}_x}}}} \right)}}{{\sin \left( { - \pi {{\bar d}_x}\frac{1}{{2{N_s}{{\bar d}_x}}}} \right)}}} \right|^2} = \frac{{{{\left| {{e^{jh\left( {{\alpha _p},{\Delta _p},{{\bar \Phi }_{_p}}} \right)}} + {e^{jh\left( {{\alpha _{p + 1}},{\Delta _p},{{\bar \Phi }_{_{p + 1}}}} \right)}}} \right|}^2}}}{{{{\sin }^2}\left( {\frac{\pi }{{2{N_s}}}} \right)}}.\label{arrayGainAlongIntersectionPoints}
\end{align}
\hrulefill 
\vspace*{4pt} 
\end{figure*}
 \begin{equation}
 \begin{aligned}
 {\alpha _{p + 1}} - {\alpha _p} &=  - 2\pi {N_s}{{\bar d}_x}\left( {{{\bar \Phi }_1} + \frac{{2p - 1}}{{2{N_s}{{\bar d}_x}}}} \right) + \frac{{\pi \left( {{N_s} - 1} \right)}}{{{N_s}}} + 2k\pi  \\
 &=  - 2\pi {N_s}{{\bar d}_x}{{\bar \Phi }_1} - \frac{\pi }{{{N_s}}} + 2k\pi.
\end{aligned}
 \end{equation}
 Let $k=0$, the common phase term for each sub-array can be given by
 \begin{equation}
 {\alpha _l} =  - \left( {2\pi {N_s}{{\bar d}_x}{{\bar \Phi }_1} + \frac{\pi }{{{N_s}}}} \right)l,\ l = 1, \cdots ,L. \label{subArrayCommonPhaseTermswithoutq}
 \end{equation}
 With the obtained sub-array common phase terms, it is observed from \eqref{arrayGainAlongIntersectionPoints} that for sufficiently large $N_s$, the resulting array gain at the intersection points is ${g_2}\left( {L,{N_s},\left\{ {{\alpha _l},{{\bar \Phi }_l}} \right\},\Delta ={\Delta _p} } \right) = \frac{4}{{{{\sin }^2}\left( {\frac{\pi }{{2{N_s}}}} \right)}} \approx \frac{{16}}{{{\pi ^2}}}N_s^2$, which is also proportional to $N_s^2$, similar to that for $\Delta  = {{\bar \Phi }_l}$. In addition, with such a design, the array gain at other spatial frequencies can be obtained numerically. Fig.~\ref{resultingBroadenedBeam} shows one example of the resulting broadened and flattened beam for a ULA with $N=512$ elements partitioned into $L=4$ sub-arrays by setting ${{\bar \Phi} _1} =  - \frac{{L - 1}}{{2{N_s}{{\bar d}_x}}}$, where the dotted line indicates the array gain of ${\rm{10}}{\log _{10}}\left( {N_s^2} \right)$~dB. It is observed that the array gain within the spatial frequency interval $\left[ {{{\bar \Phi }_1},{{\bar \Phi }_4}} \right]$ slightly fluctuates around ${\rm{10}}{\log _{10}}\left( {N_s^2} \right)$, and the maximum fluctuation is only about 1.5~dB. Thus, with the proposed beam flattening design, the array gain within the main lobe is approximated by
 \begin{equation}
 {g_2}\left( {L,{N_s},\left\{ {{{\bar \Phi }_l},{\alpha _l}} \right\},\Delta } \right) \approx N_s^2 = \frac{{{N^2}}}{{{L^2}}},{{\bar \Phi }_1} \le \Delta  \le {{\bar \Phi }_L}. \label{MainLobeArrayGain}
 \end{equation}
 Besides, Fig.~\ref{resultingBroadenedBeam} also shows that the coverage beamwidth of the flattened beam is approximately $\left[ {{{\bar \Phi }_1} - \frac{1}{{2{N_s}{{\bar d}_x}}},{{\bar \Phi }_L} + \frac{1}{{2{N_s}{{\bar d}_x}}}} \right]$, within which the worst-case array gain occurs at the boundary spatial frequencies ${{{\bar \Phi }_1} - \frac{1}{{2{N_s}{{\bar d}_x}}}}$ and ${{{\bar \Phi }_L} + \frac{1}{{2{N_s}{{\bar d}_x}}}}$. A closer look at Fig.~\ref{resultingBroadenedBeam} shows that only one
 sub-array has the most contribution to the array gain of the two spatial frequencies. Thus, for sufficiently large $N_s$, by setting $\Delta  = \frac{1}{{2{N_s}{{\bar d}_x}}}$ in \eqref{array gain}, the worst-case array gain is
 \begin{equation}
 \begin{aligned}
 g_2^{{\rm{worst}}}\left( {L,{N_s},\left\{ {{{\bar \Phi }_l},{\alpha _l}} \right\},\Delta } \right) \approx \frac{4}{{{\pi ^2}}}N_s^2 = \frac{4}{{{\pi ^2}}}\frac{{{N^2}}}{{{L^2}}}&,\\
 {{\bar \Phi }_1} - \frac{1}{{2{N_s}{{\bar d}_x}}} \le \Delta  \le {{\bar \Phi }_L} + \frac{1}{{2{N_s}{{\bar d}_x}}},& \label{worstCaseArrayGain}
 \end{aligned}
 \end{equation}
 which is inversely proportional to ${L^2}$ with given total number of elements $N$. Furthermore, for the proposed beam flattening technique with sub-array manifold using $L$ sub-arrays each with $N_s$ elements, the coverage beamwidth in terms of $L$ is given by
  \begin{equation}
  {{\bar \Delta }_{{\rm{broad}}}}(L) = L{{\bar \Delta }_{{\rm{CVBW}}}}\left( {{N_s}} \right) \approx L\frac{1}{{{N_s}{{\bar d}_x}}} = \frac{{{L^2}}}{{N{{\bar d}_x}}},\label{beamBroadeningTechniqueBeamwidth}
  \end{equation}
  where $\frac{{1}}{{N{{\bar d}_x}}}$ corresponds to the coverage beamwidth of the $N$-element full array without sub-array partition or beam broadening/flatenning. Therefore, with the proposed beam broadening and flattening technique, it is observed from \eqref{worstCaseArrayGain} and \eqref{beamBroadeningTechniqueBeamwidth} that the coverage beamwidth is broadened by a factor of $L^2$, but at the cost of the same proportional reduction of the worst-case array gain. Therefore, there exists a design trade-off for choosing the optimal number of sub-arrays $L$, so as to maximize the worst-case array gain while ensuring that the beamwidth is sufficiently large to cover the entire target area.

  Based on the above discussions, an efficient solution is proposed to solve problem~\eqref{phaseShiftOptimizationEquivalentProblem}. Specifically, for the given AIRS placement $\mathbf q$, based on \eqref{theMinimumDeviation} and \eqref{theMaximumDeviation}, we define the span of the spatial frequency deviation ${\Delta _{{\rm{span}}}}\left( {\bf{q}} \right)$ associated with $\mathcal A$ as
  \begin{equation}
  {\Delta _{{\rm{span}}}}\left( {\bf{q}} \right) \buildrel \Delta \over = {\Delta _{\max }}\left( {\bf{q}} \right) - {\Delta _{\min }}\left( {\bf{q}} \right). \label{theMaximumSpatialFrequencyDeviation}
  \end{equation}
  Intuitively, to maximize the worst-case array gain within the interval $\left[ {{\Delta _{\min }}\left( {\bf{q}} \right),{\Delta _{\max }}\left( {\bf{q}} \right)} \right]$, $L$ should be large enough so that all locations in $\mathcal A$ lie within the coverage beamwidth of the AIRS. Based on \eqref{beamBroadeningTechniqueBeamwidth}, we should have ${{\bar \Delta }_{{\rm{broad}}}}(L) \ge {\Delta _{{\rm{span}}}}\left( {\bf{q}} \right)$. On the other hand, in order to maximize the approximate worst-case array gain $\frac{4}{{{\pi ^2}}}\frac{{{N^2}}}{{{L^2}}}$, $L$ should be as small as possible. Thus, we set
  \begin{equation}
  {L^*}\left( {\bf{q}} \right) = \left\lceil {\sqrt {{\Delta _{{\rm{span}}}}\left( {\bf{q}} \right)N{{\bar d}_x}} } \right\rceil. \label{theRequiredNumberofSubArrays}
  \end{equation}
  The number of elements in each sub-array is thus given by $N_s^*\left( {\bf{q}} \right) = N/{L^*}\left( {\bf{q}} \right)$. It is observed that when ${\Delta _{{\rm{span}}}}\left( {\bf{q}} \right) \le \frac{{1}}{{N{{\bar d}_x}}}$, we have $L=1$ and ${N_s} = N$. In other words, when the target coverage area $\mathcal A$ and/or the AIRS size $N$ is relatively small, no beam flatenning/broadening is needed, and the full array architecture is sufficient to cover the entire area. On the other hand, when the coverage area $\mathcal A$ and/or the AIRS size $N$ is large, we have $L>1$ in general. Furthermore, by letting ${\rm{\bar \Phi }}_1^*\left( {\bf{q}} \right) - \frac{1}{{2N_s^*\left( {\bf{q}} \right){{\bar d}_x}}} = {\Delta _{\min }}\left( {\bf{q}} \right)$, the starting spatial frequency is thus given by $\bar \Phi _1^*\left( {\bf{q}} \right) = {\Delta _{\min }}\left( {\bf{q}} \right) + \frac{1}{{2N_s^*\left( {\bf{q}} \right){{\bar d}_x}}}$. Substituting $\bar \Phi _1^*\left( {\bf{q}} \right)$ into \eqref{sub-arrayBeamDirection} and \eqref{subArrayCommonPhaseTermswithoutq}, the sub-array beam directions and common phase terms are given by
  \begin{equation}
  \bar \Phi _l^*\left( {\bf{q}} \right) = {\Delta _{\min }}\left( {\bf{q}} \right) + \frac{{2l - 1}}{{2N_s^*\left( {\bf{q}} \right){{\bar d}_x}}},\ l = 1, \cdots ,{L^*}\left( {\bf{q}} \right),
  \label{subArrayBeamDirections}
  \end{equation}
  \begin{equation}
  \begin{aligned}
  \alpha _l^*\left( {\bf{q}} \right) =  - \left( {2\pi N_s^*\left( {\bf{q}} \right){{\bar d}_x}{\Delta _{\min }}\left( {\bf{q}} \right) + \pi  + \frac{\pi }{{N_s^*\left( {\bf{q}} \right)}}} \right)l&,\\ \;l = 1, \cdots ,{L^*}\left( {\bf{q}} \right).& \label{subArrayCommonPhaseTerms}
  \end{aligned}
  \end{equation}
   Based on the sub-array manifold, the phase shifts ${{\bm{\theta }}^*}\left( {\bf{q}} \right)$ for (P5.1) can be obtained as
  \begin{equation}
  \small
  \begin{aligned}
  \theta _{\left( {l - 1} \right){N_s^*\left( {\bf{q}} \right)} + i}^*\left( {\bf{q}} \right) &=  - \left( {2\pi N_s^*\left( {\bf{q}} \right){{\bar d}_x}{\Delta _{\min }}\left( {\bf{q}} \right) + \pi  + \frac{\pi }{{N_s^*\left( {\bf{q}} \right)}}} \right)l\\
  &- 2\pi \left( {i - 1} \right){{\bar d}_x}\left( {{\Delta _{\min }}\left( {\bf{q}} \right) + \frac{{2l - 1}}{{2N_s^*\left( {\bf{q}} \right){{\bar d}_x}}}} \right),\\
  &\ \ \ \ \ l = 1, \cdots ,{L^*}\left( {\bf{q}} \right),\ i = 1, \cdots ,N_s^*\left( {\bf{q}} \right). \label{theObtainedPhaseShifts}
  \end{aligned}
  \end{equation}
  The proposed beam broadening and flattening technique for passive beamforming design for ULA-based AIRS is summarized in the following Proposition.
  \begin{proposition} \label{Proposed Passive Beam Solution for P5.1}
  To design a passive beamforming of an $N$-element ULA-based AIRS located at $\bf{q}$ with the beamwidth matching with the size of a target area ${\mathcal A}$, the following beam broadening and flattening technique is proposed:

   1. Calculate the span of the spatial frequency deviation ${\Delta _{{\rm{span}}}}\left( {\bf{q}} \right)$ associated with $\mathcal A$ based on \eqref{theMinimumDeviation}, \eqref{theMaximumDeviation}, and \eqref{theMaximumSpatialFrequencyDeviation}.

   2. Determine the number of sub-arrays ${L^*}\left( {\bf{q}} \right)$ based on \eqref{theRequiredNumberofSubArrays}.

   3. Set the sub-array beam directions $\left\{ {\bar \Phi _l^*\left( {\bf{q}} \right)} \right\}$ and common phase terms $\left\{ {\alpha _l^*\left( {\bf{q}} \right)} \right\}$ based on \eqref{subArrayBeamDirections} and \eqref{subArrayCommonPhaseTerms}.

   4. Obtain the phase-shift solution ${{\bm{\theta }}^*}\left( {\bf{q}} \right)$ based on \eqref{theObtainedPhaseShifts}.
  \end{proposition}


  It is worth pointing out that the worst-case array gain within the beam coverage $\left[ {\bar \Phi _1^*\left( {\bf{q}} \right) - \frac{1}{{2N_s^*\left( {\bf{q}} \right){{\bar d}_x}}},\bar \Phi _L^*\left( {\bf{q}} \right) + \frac{1}{{2N_s^*\left( {\bf{q}} \right){{\bar d}_x}}}} \right]$ is given by $\frac{4}{{{\pi ^2}}}{\left( {N_s^*\left( {\bf{q}} \right)} \right)^2}$. Thus, with the proposed design, the objective value of (P5.1) is approximated as
  \begin{equation}
  \small
  \begin{aligned}
  {f_1}\left( {{\bf{q}},{{\bm{\theta }}^*}\left( {\bf{q}} \right),{\bf{w}}} \right) \approx \frac{4}{{{\pi ^2}}}\frac{{{N^2}}}{{{{\left( {{L^*}\left( {\bf{q}} \right)} \right)}^2}}} =& \frac{4}{{{\pi ^2}}}\frac{{{N^2}}}{{{{\left\lceil {\sqrt {{\Delta _{{\rm{span}}}}\left( {\bf{q}} \right)N{{\bar d}_x}} } \right\rceil }^2}}},\\
  &\forall {\bf{w}} \in \mathcal A.\label{beamBroadeningTechniqueArrayGain}
  \end{aligned}
  \end{equation}

  It is observed that when the number of reflecting elements $N$ and/or the spatial frequency span ${{\Delta _{{\rm{span}}}}\left( {\bf{q}} \right)}$ is small such that ${\Delta _{{\rm{span}}}}\left( {\bf{q}} \right)N{{\bar d}_x} < 1$, the beamwidth of the full array can directly match the size of the target coverage area, and \eqref{beamBroadeningTechniqueArrayGain} reduces to
  \begin{equation}
  {f_1}\left( {{\bf{q}},{{\bm{\theta }}^*}\left( {\bf{q}} \right),{\bf{w}}} \right) \approx \frac{4}{{{\pi ^2}}}{N^2},\ \forall {\bf{w}} \in \mathcal A. \label{smallNumberAndSpanArrayGain}
  \end{equation}
  This implies that the worst-case array gain of all the locations in the target area quadratically increases with $N$. This result is in accordance with the single-location SNR maximization problem with the full array (see \eqref{single-location SNR}).

  \begin{figure}[!t]
  \centering
  \centerline{\includegraphics[width=3.5in,height=2.625in]{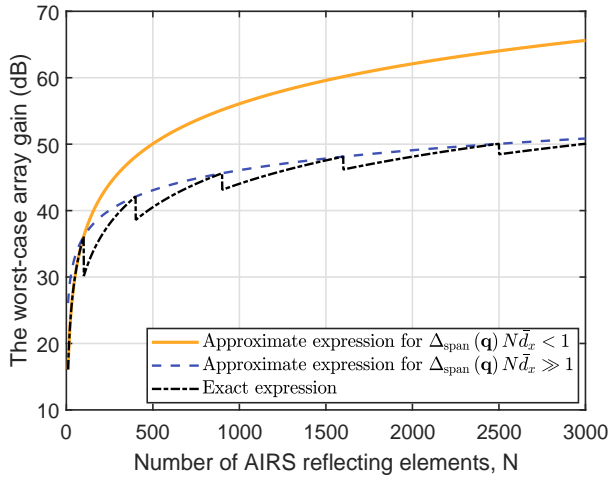}}
  \caption{The worst-case array gain versus the number of AIRS reflecting elements for area coverage.}
  \label{TheworstcaseArrayGainVersustheNumberOfAIRSReflectingElements}
  \end{figure}
  On the other hand, when $N$ and/or ${{\Delta _{{\rm{span}}}}\left( {\bf{q}} \right)}$ is sufficiently large such that ${\Delta _{{\rm{span}}}}\left( {\bf{q}} \right)N{{\bar d}_x} \gg 1$, the ceiling operator in \eqref{beamBroadeningTechniqueArrayGain} is negligible, and we have
  \begin{equation}
  \small
  {f_1}\left( {{\bf{q}},{{\bm{\theta }}^*}\left( {\bf{q}} \right),{\bf{w}}} \right) \approx  \frac{4}{{{\pi ^2}}}\frac{{{N^2}}}{{{\Delta _{{\rm{span}}}}\left( {\bf{q}} \right)N{{\bar d}_x}}} = \frac{4}{{{\pi ^2}}}\frac{N}{{{\Delta _{{\rm{span}}}}\left( {\bf{q}} \right){{\bar d}_x}}},\forall {\bf{w}} \in \mathcal A.\label{approximateArrayGain}
  \end{equation}
  The result in \eqref{approximateArrayGain} shows that for the area coverage setup with relatively wide target area size and/or large AIRS elements number $N$, the worst-case array gain over all the locations in the target area increases linearly with $N$, which is in contrast to \eqref{single-location SNR} or \eqref{smallNumberAndSpanArrayGain}. This is expected since for area coverage with relatively wide area and/or large $N$, the $N$ AIRS elements need to be partitioned into sub-arrays to form broadened and flattened beams, at the cost of reduced array gain along the main beam. For a fixed spatial frequency span ${\Delta _{{\rm{span}}}}\left( {\bf{q}} \right) = 0.1$, Fig.~\ref{TheworstcaseArrayGainVersustheNumberOfAIRSReflectingElements} plots the worst-case array gain \eqref{beamBroadeningTechniqueArrayGain} versus $N$, together with its approximations \eqref{smallNumberAndSpanArrayGain} and \eqref{approximateArrayGain}. It is observed that \eqref{smallNumberAndSpanArrayGain} and \eqref{approximateArrayGain} well approximate \eqref{beamBroadeningTechniqueArrayGain} at the small and large $N$ regimes, respectively, with the worst-case SNR of the target area first increasing quadratically with $N$, and then only linearly when $N$ becomes large.

\subsubsection{AIRS Placement Optimization} \label{ULAAIRSPlacementOptimization}
With the proposed passive beamforming design in Proposition~\ref{Proposed Passive Beam Solution for P5.1} and the corresponding worst-case array gain in \eqref{beamBroadeningTechniqueArrayGain}, the AIRS placement optimization problem (P5.2) is greatly simplified. Specifically, by substituting \eqref{beamBroadeningTechniqueArrayGain} into (P5.2), we have
\begin{equation}
\small
 \mathop {\max }\limits_{\bf{q}} {\rm{ }}\mathop {\min }\limits_{{\bf{w}} \in {\cal A}} {\rm{  }}\frac{4}{{{\pi ^2}}}\frac{{{N^2}}}{{{{\left\lceil {\sqrt {{\Delta _{{\rm{span}}}}\left( {\bf{q}} \right)N{{\bar d}_x}} } \right\rceil }^2}\left( {{H^{\rm{2}}}{\rm{ + }}{{\left\| {{\bf{q}} - {\bf{w}}} \right\|}^2}} \right)\left( {{H^{\rm{2}}}{\rm{ + }}{{\left\| {\bf{q}} \right\|}^2}} \right)}}.\label{AIRSPlacementOptimization1}
\end{equation}
Since only the term ${{{\left\| {{\bf{q}} - {\bf{w}}} \right\|}^2}}$ in \eqref{AIRSPlacementOptimization1} depends on $\bf{w}$, by letting ${d_{\max }}\left( {\bf{q}} \right) \buildrel \Delta \over = \mathop {\max }\limits_{{\bf{w}} \in A} \left\| {{\bf{q}} - {\bf{w}}} \right\|$, problem \eqref{AIRSPlacementOptimization1} can be equivalently transformed to
\begin{equation}
\mathop {\min }\limits_{\bf{q}} {\rm{  }}{\left\lceil {\sqrt {{\Delta _{{\rm{span}}}}\left( {\bf{q}} \right)N{{\bar d}_x}} } \right\rceil ^2}\left( {{H^{\rm{2}}}{\rm{ + }}d_{\max }^2\left( {\bf{q}} \right)} \right)\left( {{H^{\rm{2}}}{\rm{ + }}{{\left\| {\bf{q}} \right\|}^2}} \right).\label{AIRSPlacementOptimization3}
\end{equation}
Problem \eqref{AIRSPlacementOptimization3} shows that the optimal AIRS placement should achieve a balance between minimizing the spatial frequency (or angular) span associated with the target coverage area $\mathcal A$, as well as minimizing the cascaded path loss from the source node to the worst-case destination node. Note that if the target area $\mathcal A$ is sufficiently small so that the first term in \eqref{AIRSPlacementOptimization3} is always 1 for all $\bf{q}$, then problem \eqref{AIRSPlacementOptimization3} degrades to (P4). Therefore, the AIRS placement optimization in \eqref{AIRSPlacementOptimization3} can be regarded as a generalization of that for the single-location SNR maximization. Meanwhile, it is observed from \eqref{AIRSPlacementOptimization3} that for a given AIRS placement $\bf{q}$, the worst-case SNR in $\mathcal A$
occurs at the location that is farthest from the AIRS placement, i.e., $\arg \mathop {\max }\limits_{{\bf{w}} \in {\cal A}} {\left\| {{\bf{q}} - {\bf{w}}} \right\|^2}$, which should be one of the boundary points in $\mathcal A$.

Furthermore, since the target area $\mathcal A$ is symmetric about the $x$-axis, it can be shown that the optimal $\bf q$ to \eqref{AIRSPlacementOptimization3} should lie in the $x$-axis, i.e., we should have ${{\bf{q}}} = {\left[ {{q_x},0} \right]^T}$. Thus, \eqref{AIRSPlacementOptimization3} reduces to the univariate optimization problem, i.e.,
\begin{equation}
\mathop {\min }\limits_{{q_x}} {\left\lceil {\sqrt {{\Delta _{{\rm{span}}}}\left( {{q_x}} \right)N{{\bar d}_x}} } \right\rceil ^2}\left( {{H^{\rm{2}}}{\rm{ + }}d_{\max }^2\left( {{q_x}} \right)} \right)\left( {{H^{\rm{2}}}{\rm{ + }}{{\left| {{q_x}} \right|}^2}} \right). \label{AIRSPlacementOptimization4}
\end{equation}
It can be observed that starting from ${q_x} = 0$, as $q_x$ increases so that the AIRS moves closer to the target area, the first term and the third term in \eqref{AIRSPlacementOptimization4} increase, while the second term decreases. As a result, there in general exists a non-trivial optimal AIRS placement $q_x^*$ to balance the above three terms, which can be efficiently found via a one-dimensional search. As such, the AIRS placement problem (P5.2) and hence (P5) are solved. The main procedures for solving (P5.2) are summarized in Algorithm~\ref{alg2}.
\begin{algorithm}[H]  
\caption{AIRS Placement Optimization for $(\rm P5.2)$}
\label{alg2}
\begin{algorithmic}[1]
\STATE  \textbf{Initialization:} Set the search regime of AIRS placement ${q_x} \in \left[ {{q_{\min }}, {q_{\max }}} \right]$.
\STATE For all candidate placement $q_x$, obtain the maximum distance ${d_{\max }}\left( q_x \right)$, and the spatial frequency span based on \eqref{theMaximumSpatialFrequencyDeviation}. Calculate the cost value in \eqref{AIRSPlacementOptimization4}.
\STATE Choose $q_x$ that gives the minimum cost value in \eqref{AIRSPlacementOptimization4} as the optimal placement, and return the corresponding passive beamforming based on Proposition~\ref{Proposed Passive Beam Solution for P5.1}.
\end{algorithmic}
\end{algorithm}

\subsection{The General Case of UPA-Based AIRS} \label{generalCaseofUPA-BasedAIRS}
Based on the results obtained for ULA-based AIRS, in this subsection, we extend our proposed solution to the general case of UPA-based AIRS by solving problem (P2) with the general SNR expression \eqref{SNR at location w}. Different from the ULA-based AIRS, the UPA-based AIRS is able to achieve passive beam steering over both spatial frequency dimensions $\bar \Phi$ and $\bar \Omega$, as evident from \eqref{UPAReceiveArrayResponseAtAIRS}, \eqref{UPATransmitArrayResponseAtAIRS} and \eqref{SNR at location w}. Therefore, the difficulty in solving problem (P2) for UPA-based AIRS lies in that a phase shift ${\theta _{{n_x},{n_y}}}$ by the $(n_x,n_y)$th passive element would in general impact the beam steering for both $\bar \Phi$ and $\bar \Omega$ dimensions (see \eqref{SNR at location w}). Fortunately, motivated by the result obtained in the previous subsection, we are able to treat the UPA-based AIRS as two {\it decoupled} ULA-based AIRS along the $x$- and $y$-axis, respectively.

The key to achieve such a decoupling is to restrict the phase shift design $\theta _{{n_x},{n_y}}$ as a sum of two phase shifts, i.e., ${\theta _{{n_x},{n_y}}} = {\theta _{{n_x}}} + {\theta _{{n_y}}}$, $1\leq n_x\leq N_x$, $1\leq n_y \leq N_y$. As such, the resulting SNR in \eqref{SNR at location w} can be re-expressed as \eqref{UPASNRatLocationW}, shown at the top of this page. Note that such a simplification is sub-optimal in general, since it narrows down the optimization variable space of problem (P2), where the number of independent phase-shift variables to be optimized is reduced from ${N_x}{N_y}$ to ${N_x} + {N_y}$. However, it serves as a useful technique for efficient 3D passive beamforming design in practice. A closer look at the numerator of \eqref{UPASNRatLocationW} reveals that the resulting SNR can be treated as that determined by the array gains achieved by two independent ULAs, along $x$- and $y$-axis, respectively. Therefore, by applying the similar phase-shift design in Proposition~\ref{Proposed Passive Beam Solution for P5.1} along the $x$- and $y$-axis for beam steering over spatial frequencies $\bar \Phi$ and $\bar \Omega$, respectively, a 3D broadened and flattened beam pattern can be achieved with its beamwidth matching the size of the target coverage area.

Specifically, for any given AIRS placement $\bf q$, denote by ${{\Delta _{{\rm{span}},x}}\left( {\bf{q}} \right)}$ and ${{\Delta _{{\rm{span}},y}}\left( {\bf{q}} \right)}$ the span of the spatial frequency deviation associated with the target area $\mathcal A$ along the $x$- and $y$-axis, respectively, where ${{\Delta _{{\rm{span}},x}}\left( {\bf{q}} \right)}$ can be obtained based on \eqref{theMinimumDeviation}, \eqref{theMaximumDeviation} and \eqref{theMaximumSpatialFrequencyDeviation}, and ${{\Delta _{{\rm{span}},y}}\left( {\bf{q}} \right)}$ can be similarly obtained as
\newcounter{mytempeqncnt4}
\begin{figure*}
\normalsize
\setcounter{mytempeqncnt4}{\value{equation}}
\begin{equation}
 \gamma \left( {{\bf{q}},{\bf{\Theta }},{\bf{w}}} \right) = \frac{{\bar P\beta _0^2M{{\left| {\sum\limits_{{n_x} = 1}^{{N_x}} {{e^{j\left( {{\theta _{{n_x}}} + 2\pi \left( {{n_x} - 1} \right){{\bar d}_x}\left[ {{{\bar \Phi }_T}\left( {{\bf{q}},{\bf{w}}} \right) - {{\bar \Phi }_R}\left( {\bf{q}} \right)} \right]} \right)}}} } \right|}^2}{{\left| {\sum\limits_{{n_y} = 1}^{{N_y}} {{e^{j\left( {{\theta _{{n_y}}} + 2\pi \left( {{n_y} - 1} \right){{\bar d}_y}\left[ {{{\bar \Omega }_T}\left( {{\bf{q}},{\bf{w}}} \right) - {{\bar \Omega }_R}\left( {\bf{q}} \right)} \right]} \right)}}} } \right|}^2}}}{{\left( {{H^2} + {{\left\| {{\bf{q}} - {\bf{w}}} \right\|}^2}} \right)\left( {{H^2} + {{\left\| {\bf{q}} \right\|}^2}} \right)}}.\label{UPASNRatLocationW}
\end{equation}
\hrulefill 
\vspace*{-0.5cm} 
\end{figure*}
 \begin{equation}
 \begin{aligned}
 {\Delta _{{\rm{span}},y}}\left( {\bf{q}} \right) &= {\Delta _{{\rm{max}},y}}\left( {\bf{q}} \right) - {\Delta _{{\rm{min}},y}}\left( {\bf{q}} \right)\\
 &= \mathop {\max }\limits_{{\bf{w}} \in {\rm{{\cal A}}}} {{\bar \Omega} _T}\left( {{\bf{q}},{\bf{w}}} \right) - \mathop {\min }\limits_{{\bf{w}} \in {\rm{{\cal A}}}} {{\bar \Omega} _T}\left( {{\bf{q}},{\bf{w}}} \right),\label{theMaximumSpatialFrequencyAlongy-axisDeviation}
 \end{aligned}
 \end{equation}
 where ${\Delta _{{\rm{max}},y}}\left( {\bf{q}} \right)$ and ${\Delta _{{\rm{min}},y}}\left( {\bf{q}} \right)$ are the maximum and minimum deviation of ${{{\bar \Omega} _T}\left( {{\bf{q}},{\bf{w}}} \right)}$ from ${{{\bar \Omega} _R}\left( {\bf{q}} \right)}$, respectively. Similar to \eqref{theRequiredNumberofSubArrays}, to ensure that the coverage beamwidth of the two decoupled beams match the size of the target coverage area, the required number of sub-arrays along the $x$- and $y$-axis are given by
 \begin{equation}
 L_a^{\rm{*}}\left( {\bf{q}} \right) = \left\lceil {\sqrt {{\Delta _{{\rm{span}},a}}\left( {\bf{q}} \right){N_a}{{\bar d}_a}} } \right\rceil,\ a \in \left\{ {x,y} \right\}. \label{UPAtheRequiredNumberofSubArrays}
 \end{equation}
 With the obtained number of sub-arrays, the sub-array beam direction, denoted as $\bar S_{l,a}^*\left( {\bf{q}} \right)$, can be determined based on \eqref{subArrayBeamDirections}, i.e.,
  \begin{equation}
  \bar S_{l,a}^*\left( {\bf{q}} \right) = {\Delta _{\min ,a}}\left( {\bf{q}} \right) + \frac{{2l - 1}}{{2N_{s,a}^*\left( {\bf{q}} \right){{\bar d}_a}}},\ l = 1, \cdots ,L_a^{\rm{*}}\left( {\bf{q}} \right),  \label{UPAsubArrayBeamDirections}
  \end{equation}
  where $N_{s,a}^*\left( {\bf{q}} \right) = {N_a}/L_a^{\rm{*}}\left( {\bf{q}} \right)$ denotes the number of elements in each sub-array along the $a$-axis, $a \in \left\{ {x,y} \right\}$. Similarly, let $\alpha _{l.a}^*\left( {\bf{q}} \right),a \in \left\{ {x,y} \right\}$ denote the sub-array common phase terms, which can be obtained as
  \begin{equation}
  \begin{aligned}
  \alpha _{l.a}^*\left( {\bf{q}} \right) =  - \left( {2\pi N_{s,a}^*\left( {\bf{q}} \right){{\bar d}_a}{\Delta _{\min ,a}}\left( {\bf{q}} \right) + \pi  + \frac{\pi }{{N_{s,a}^*\left( {\bf{q}} \right)}}} \right)l&,\\
  l = 1, \cdots ,L_a^{\rm{*}}\left( {\bf{q}} \right).\ \ \ \ \ \ & \label{UPAsubArrayCommonPhaseTermsAlong}
  \end{aligned}
  \end{equation}
  Furthermore, based on \eqref{theObtainedPhaseShifts}, the phase shifts $\theta _{{n_a}}^*\left( {\bf{q}} \right)$ along the $a$-axis can be obtained as \eqref{UPAtheObtainedPhaseShifts}, shown at the top of the next page.
  \newcounter{mytempeqncnt5}
\begin{figure*}
\normalsize
\setcounter{mytempeqncnt5}{\value{equation}}
\begin{align}
  \theta _{\left( {l - 1} \right)N_{s,a}^*\left( {\bf{q}} \right) + i}^*\left( {\bf{q}} \right) =  - \left( {2\pi N_{s,a}^*\left( {\bf{q}} \right){{\bar d}_a}{\Delta _{\min ,a}}\left( {\bf{q}} \right) + \pi  + \frac{\pi }{{N_{s,a}^*\left( {\bf{q}} \right)}}} \right)l- 2\pi \left( {i - 1} \right){{\bar d}_a}\left( {{\Delta _{\min ,a}}\left( {\bf{q}} \right) + \frac{{2l - 1}}{{2N_{s,a}^*\left( {\bf{q}} \right){{\bar d}_a}}}} \right),\notag\\
  l = 1, \cdots ,L_a^{\rm{*}}\left( {\bf{q}} \right),i = 1, \cdots ,N_{s,a}^*\left( {\bf{q}} \right). \label{UPAtheObtainedPhaseShifts}
\end{align}
\vspace*{-0.5cm}
\end{figure*}

Therefore, the phase shifts for UPA-based AIRS can be obtained accordingly as $\theta _{{n_x},{n_y}}^*\left( {\bf{q}} \right) = \theta _{{n_x}}^*\left( {\bf{q}} \right) + \theta _{{n_y}}^*\left( {\bf{q}} \right),1 \le {n_x} \le {N_x},1 \le {n_y} \le {N_y}$. Based on the above discussions, for any given AIRS placement $\bf q$, the 3D passive beamforming based on the proposed beam broadening and flattening technique is summarized as Algorithm~\ref{alg3}.
\begin{algorithm}[h]  
\caption{3D Passive Beam Design for UPA-based AIRS}
\label{alg3}
\begin{algorithmic}[1]
\STATE  \textbf{Input:} The AIRS placement $\bf{q}$, and the target coverage area $\mathcal A$.
\STATE  Calculate the span of the spatial frequency deviations ${\Delta _{{\rm{span}},x}}\left( {\bf{q}} \right)$ and ${\Delta _{{\rm{span}},y}}\left( {\bf{q}} \right)$
based on \eqref{theMaximumSpatialFrequencyDeviation} and \eqref{theMaximumSpatialFrequencyAlongy-axisDeviation}, respectively.
\STATE  Determine the required number of sub-arrays $L_x$ and $L_y$ based on \eqref{UPAtheRequiredNumberofSubArrays}.
\STATE  Set the sub-array beam directions $\left\{ {\bar S_{l,x}^*\left( {\bf{q}} \right)} \right\}$ and $\left\{ {\bar S_{l,y}^*\left( {\bf{q}} \right)} \right\}$ based on \eqref{UPAsubArrayBeamDirections}, and common phase terms $\left\{ {\alpha _{l,x}^*\left( {\bf{q}} \right)} \right\}$, $\left\{ {\alpha _{l,y}^*\left( {\bf{q}} \right)} \right\}$ based on \eqref{UPAsubArrayCommonPhaseTermsAlong}.  \\
\STATE  Obtain the phase shifts $\theta _{{n_x}}^*\left( {\bf{q}} \right)$ and $\theta _{{n_y}}^*\left( {\bf{q}} \right)$ based on \eqref{UPAtheObtainedPhaseShifts}.
\STATE \textbf{Output:} The phase shifts $\theta _{{n_x},{n_y}}^*\left( {\bf{q}} \right) = \theta _{{n_x}}^*\left( {\bf{q}} \right) + \theta _{{n_y}}^*\left( {\bf{q}} \right)$.
\end{algorithmic}
\end{algorithm}

 Similar to \eqref{beamBroadeningTechniqueArrayGain}, the array gain for the entire area $\mathcal A$ can be further characterized by \eqref{UPAapproximateArrayGain}, shown at the top of the next page.

 Based on \eqref{UPAapproximateArrayGain}, the AIRS placement optimization problem can be similarly solved as that in Section~\ref{ULAAIRSPlacementOptimization}. Specifically, similar to \eqref{AIRSPlacementOptimization3}, the problem can be first equivalently transformed to \eqref{UPAAIRSPlacementOptimization}, shown at the top of the next page. It is observed from  \eqref{UPAAIRSPlacementOptimization} that the placement optimization for UPA-based AIRS with 3D passive beamforming needs to strike a balance between minimizing the angular spans along both $x$- and $y$-axis, as well as minimizing the cascaded path loss. Furthermore, as the target area $\mathcal A$ is symmetric over the $x$-axis, so is the optimized AIRS placement, i.e., $q_y^* =  - \frac{{{N_y}{d_y}}}{2}$. In addition, similar to Algorithm~\ref{alg2}, the optimized placement $q_x^*$ can also be found via a one-dimensional search. Thus, an efficient solution is obtained for (P2) via the joint placement and 3D beamforming design for the general UPA-based AIRS.
\newcounter{mytempeqncnt6}
\begin{figure*}
\normalsize
\setcounter{mytempeqncnt6}{\value{equation}}
\begin{equation}
{f_1}\left( {{\bf{q}},{{\bm{\theta }}^*}\left( {\bf{q}} \right),{\bf{w}}} \right) \approx \frac{{16}}{{{\pi ^4}}}\frac{{N_x^2}}{{{{\left\lceil {\sqrt {{\Delta _{{\rm{span}},x}}\left( {\bf{q}} \right){N_x}{{\bar d}_x}} } \right\rceil }^2}}}\frac{{N_y^2}}{{{{\left\lceil {\sqrt {{\Delta _{{\rm{span}},y}}\left( {\bf{q}} \right){N_y}{{\bar d}_y}} } \right\rceil }^2}}},\ \forall {\bf{w}} \in \mathcal A. \label{UPAapproximateArrayGain}
\end{equation}
\begin{equation}
\mathop {\min }\limits_{\bf{q}} {\left\lceil {\sqrt {{\Delta _{{\rm{span}},x}}\left( {\bf{q}} \right){N_x}{{\bar d}_x}} } \right\rceil ^2}{\left\lceil {\sqrt {{\Delta _{{\rm{span}},y}}\left( {\bf{q}} \right){N_y}{{\bar d}_y}} } \right\rceil ^2}\left( {{H^{\rm{2}}}{\rm{ + }}d_{\max }^2\left( {\bf{q}} \right)} \right)\left( {{H^{\rm{2}}}{\rm{ + }}{{\left\| {\bf{q}} \right\|}^2}} \right). \label{UPAAIRSPlacementOptimization}
\end{equation}
\hrulefill 
\vspace*{4pt} 
\end{figure*}

\section{Numerical Results}
 In this section, numerical results are provided to evaluate the performance of our proposed design. The altitude of the AIRS is set as $H=100$~m. Unless otherwise stated, the noise and transmit power are set as ${\sigma ^2} =  - 110$~dBm and $P=20$~dBm, respectively, and the reference channel power gain is ${\beta _0} =  - 40$~dB, corresponding to a carrier frequency of 2.4~GHz. The number of transmit antennas at the source node is $M=64$. The separation of adjacent elements along the $x$- and $y$-axis are $d_x = d_y = \lambda /10$ .

\begin{figure}[!t]
  \centering
  \centerline{\includegraphics[width=3.5in,height=2.625in]{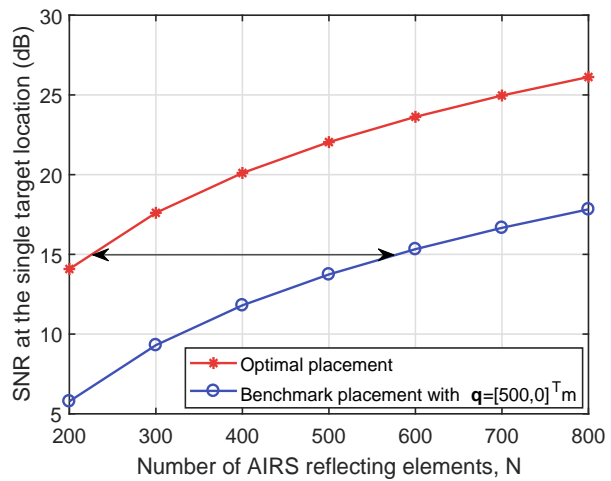}}
  \caption{SNR versus the number of AIRS reflecting elements for single-location SNR maximization.}
  \label{SNRAtTheSingleLocationVersusTheNumberofReflectingElements}
  \end{figure}

 We first consider the single-location SNR maximization problem studied in Section~\ref{SingleLocationSNREnhancementCase}. Fig.~\ref{SNRAtTheSingleLocationVersusTheNumberofReflectingElements} shows the achievable SNR for the target destination location at $[1000,0]^T$~m versus the number of AIRS reflecting elements, $N$. For the considered setup, it follows from Proposition~\ref{PropositionoptimalPlacementsinglelocation} that $\rho=10$, and the optimal AIRS deployment coefficients are ${\xi ^*}\left( \rho  \right) =  0.0101$ and $0.9899$, i.e., the AIRS should be placed either close to the source node or to the target location. As a comparison, we also consider a benchmark AIRS placement scheme with the AIRS simply placed above the midpoint between the source and the target location, i.e., ${\bf{q}} = {\left[ {500,0} \right]^T}$~m. For both the proposed optimal placement and benchmark placement, the optimal passive beamforming with phase shifts given in~\eqref{single location optimal phase shift} is applied at the AIRS to achieve coherent reflected signal combination at the target location. It is observed that for both placement schemes, the optimal SNR increases quadratically with the number of AIRS elements, as expected. In addition, the optimal placement significantly outperforms the benchmark placement. For example, to achieve a target SNR of 15~dB, the number of AIRS elements required for the benchmark placement is about 580, while this number is significantly reduced to about 225 for the proposed optimal placement. This example demonstrates the importance of exploiting the flexible AIRS placement for communication performance optimization.

\begin{figure*}
\centering
\hspace{-0.5cm}
\subfigure[$x_l=250$~m, $x_u=750$~m]{
\begin{minipage}[t]{0.32\textwidth}
\centering
\centerline{\includegraphics[width=2.4in,height=1.8in]{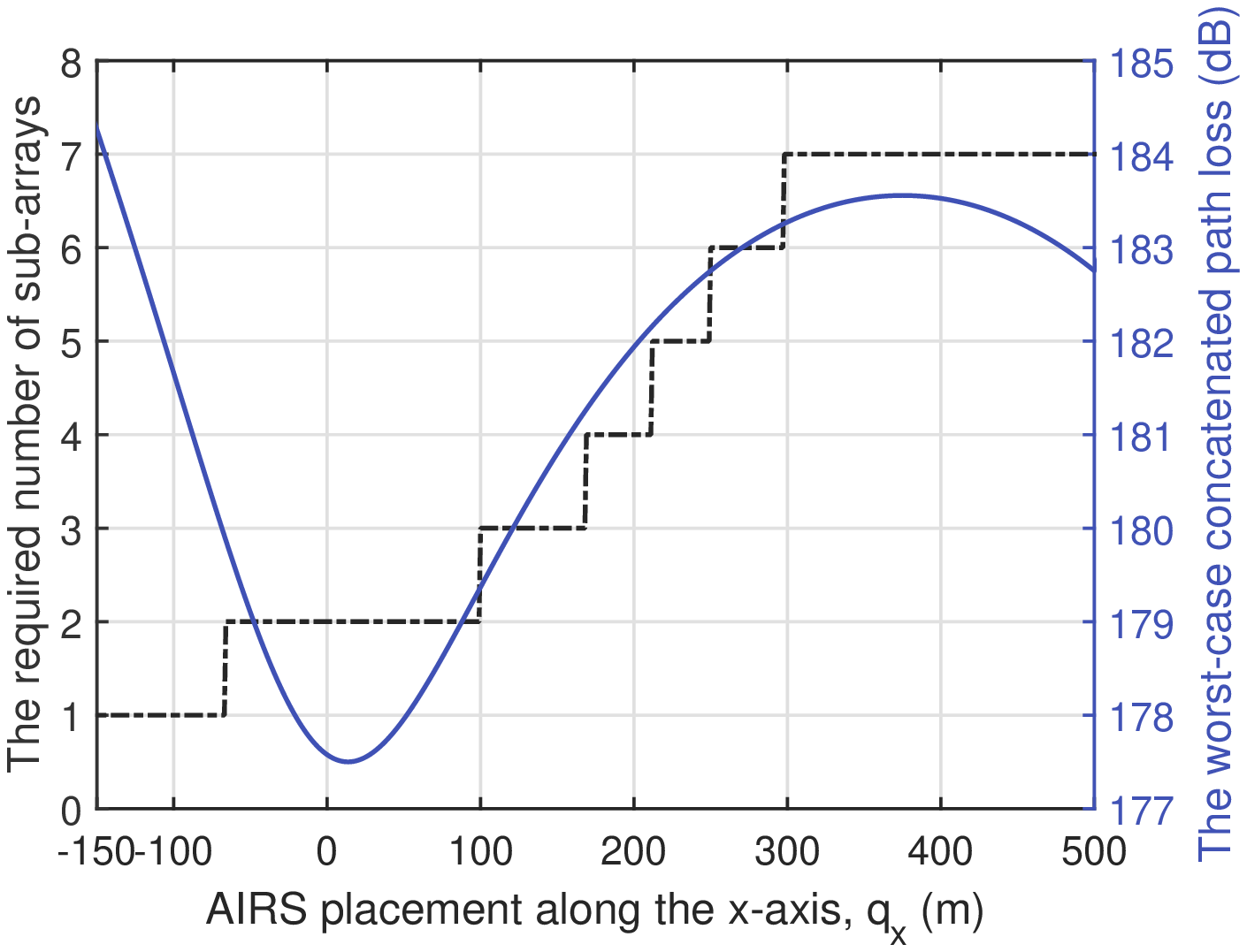}}
\vspace{0.3cm}
\end{minipage}
}
\subfigure[$x_l=500$~m, $x_u=1500$~m]{
\begin{minipage}[t]{0.32\textwidth}
\centering
\centerline{\includegraphics[width=2.4in,height=1.8in]{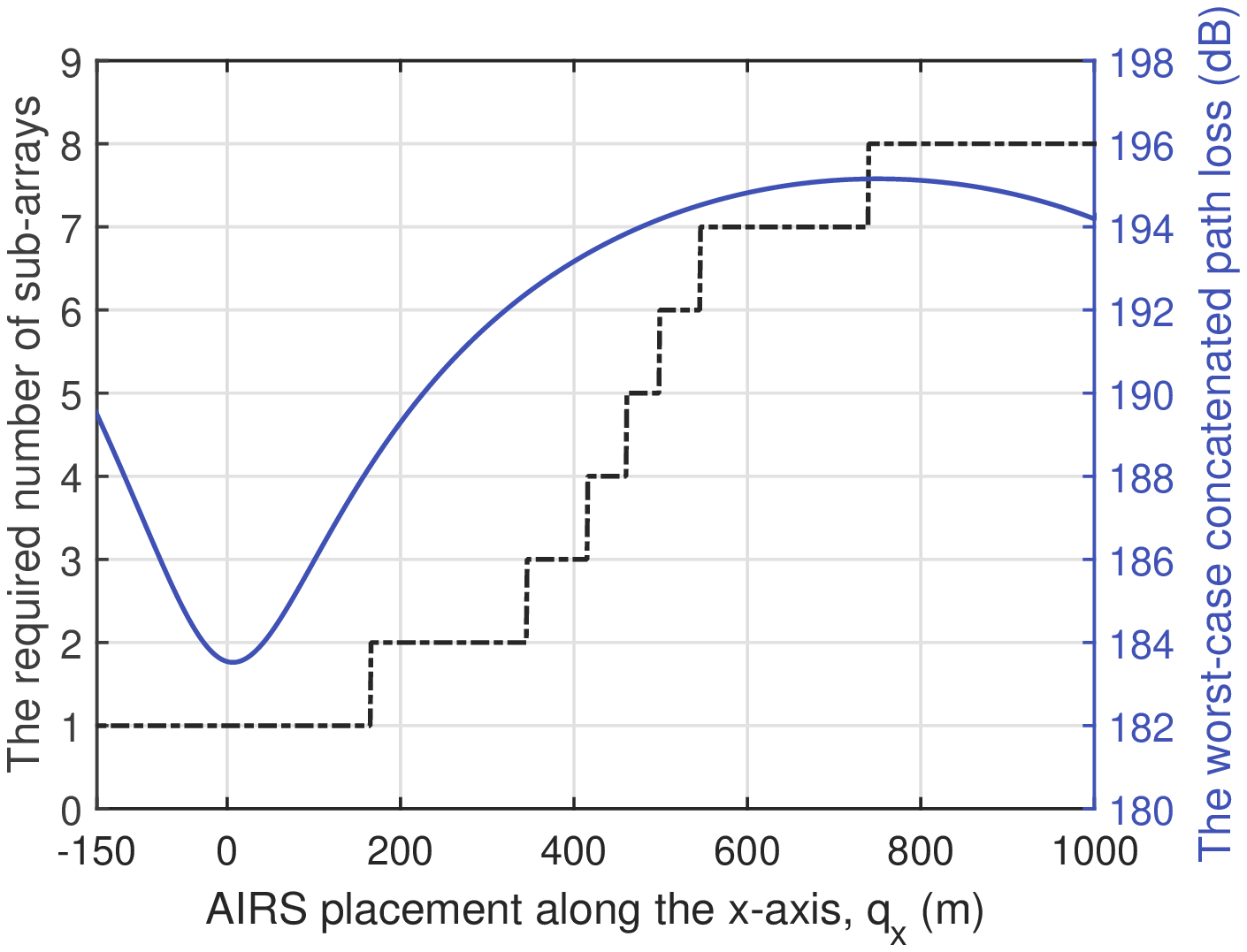}}
\vspace{0.3cm}
\end{minipage}
}
\subfigure[$x_l=155$~m, $x_u=325$~m]{
\begin{minipage}[t]{0.32\textwidth}
\centering
\centerline{\includegraphics[width=2.4in,height=1.8in]{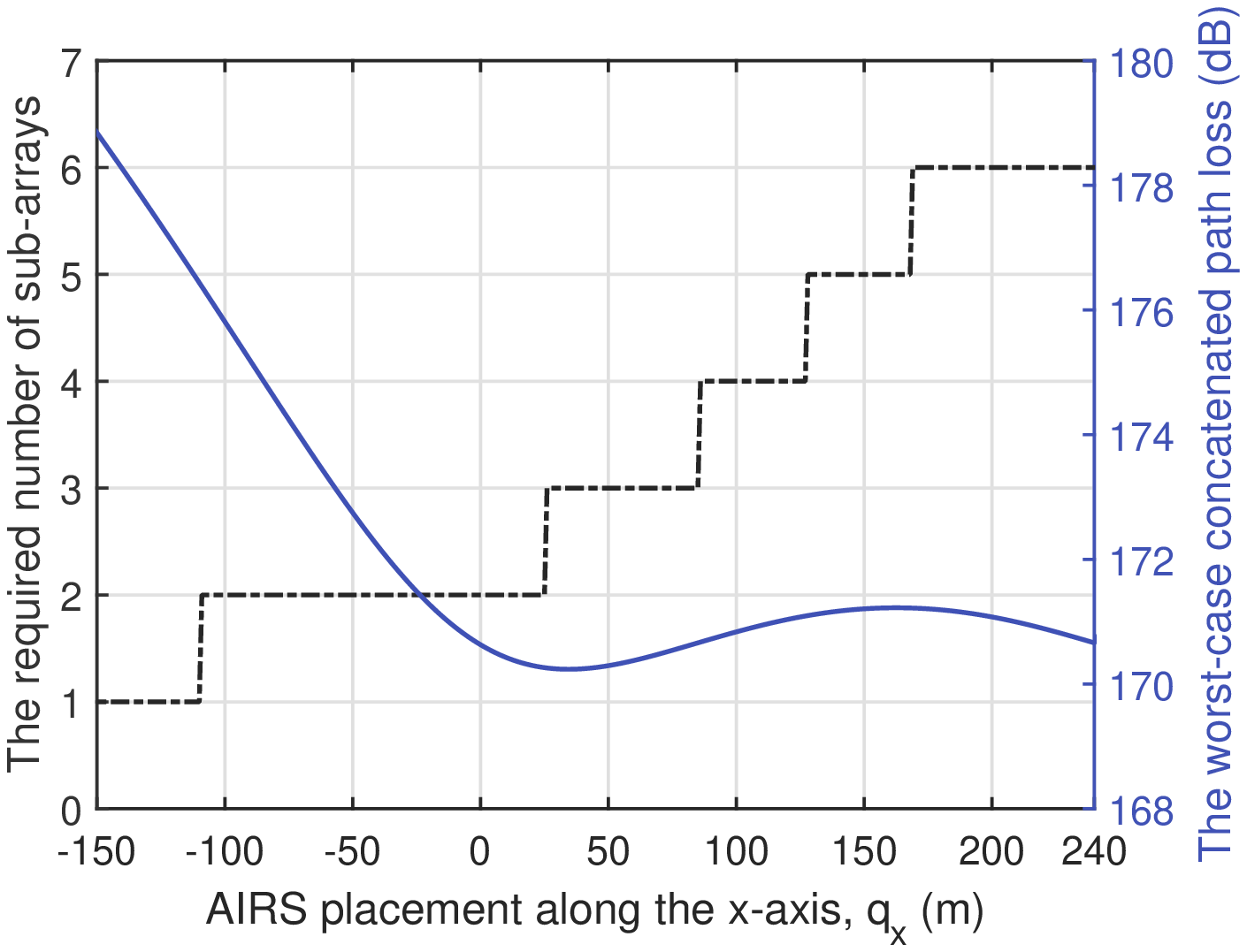}}
\vspace{0.3cm}
\end{minipage}
}
\caption{The required number of sub-arrays and the worst-case concatenated path loss versus the AIRS placement along the $x$-axis for 1D coverage with the ULA-based AIRS.}
\label{LossandNumberLineSegmentCase}
\end{figure*}

\begin{figure*}
\centering
\hspace{-0.5cm}
\subfigure[$x_l=250$~m, $x_u=750$~m]{
\begin{minipage}[t]{0.32\textwidth}
\centering
\centerline{\includegraphics[width=2.4in,height=1.8in]{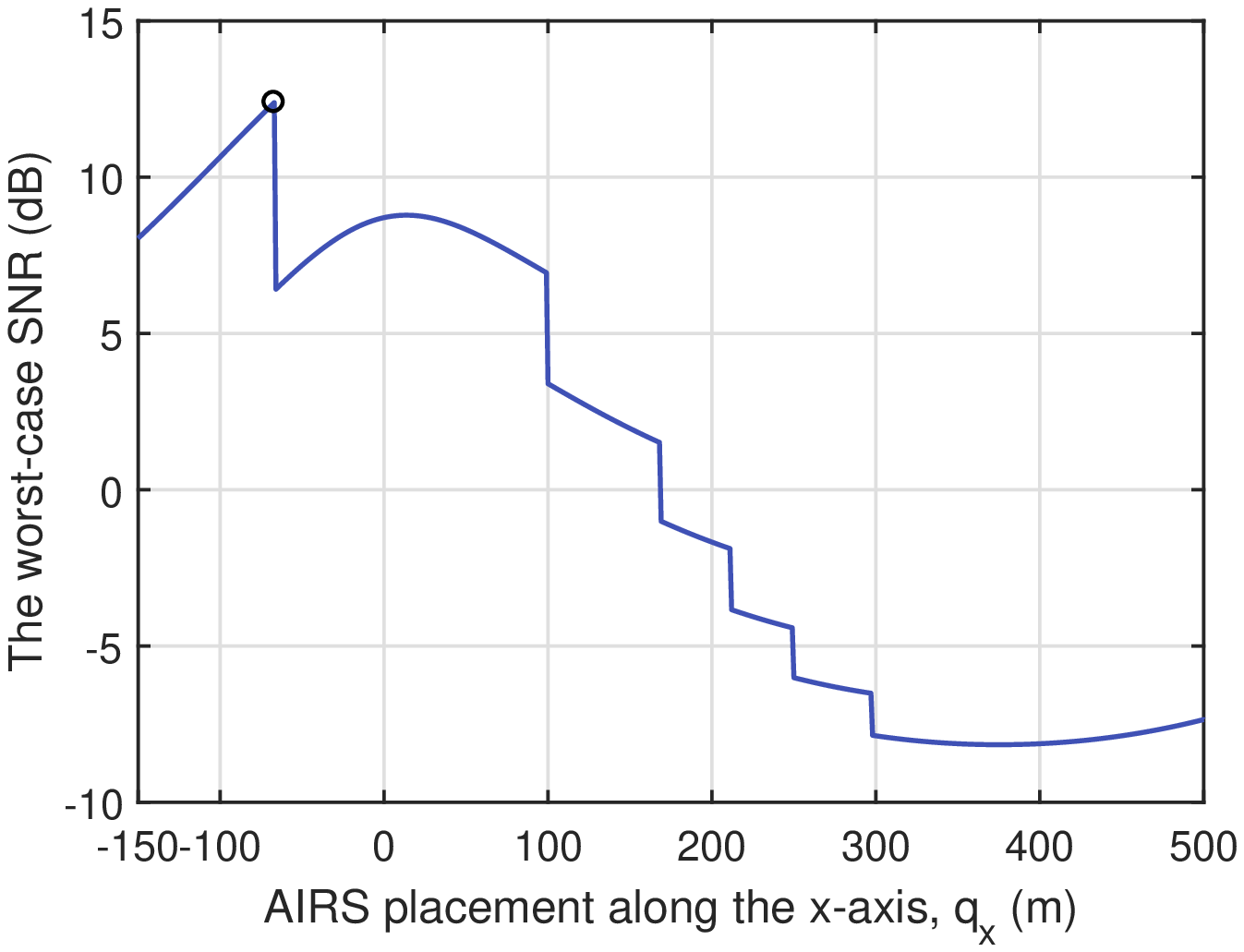}}
\vspace{0.3cm}
\end{minipage}
}
\subfigure[$x_l=500$~m, $x_u=1500$~m]{
\begin{minipage}[t]{0.32\textwidth}
\centering
\centerline{\includegraphics[width=2.4in,height=1.8in]{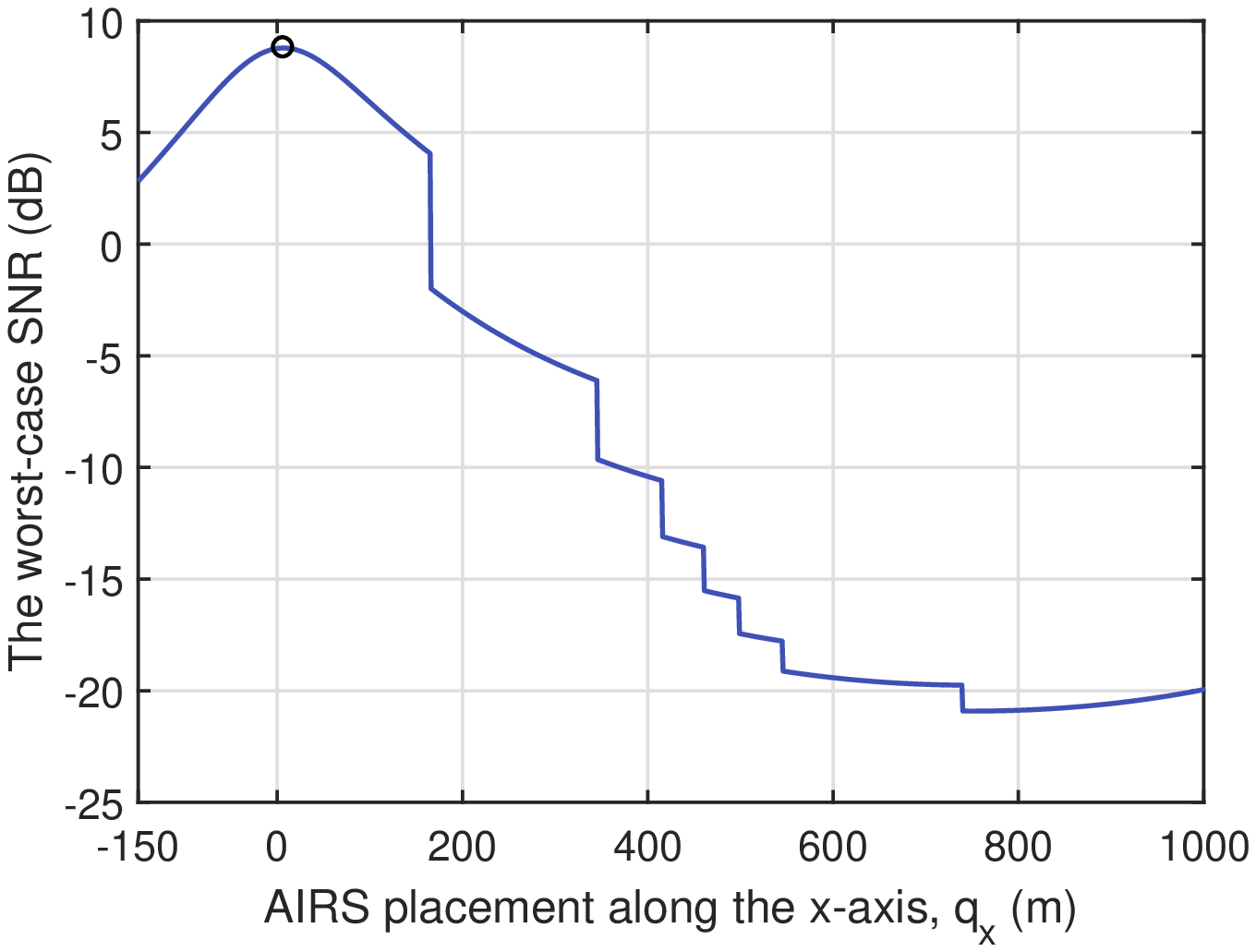}}
\vspace{0.3cm}
\end{minipage}
}
\subfigure[$x_l=155$~m, $x_u=325$~m]{
\begin{minipage}[t]{0.32\textwidth}
\centering
\centerline{\includegraphics[width=2.4in,height=1.8in]{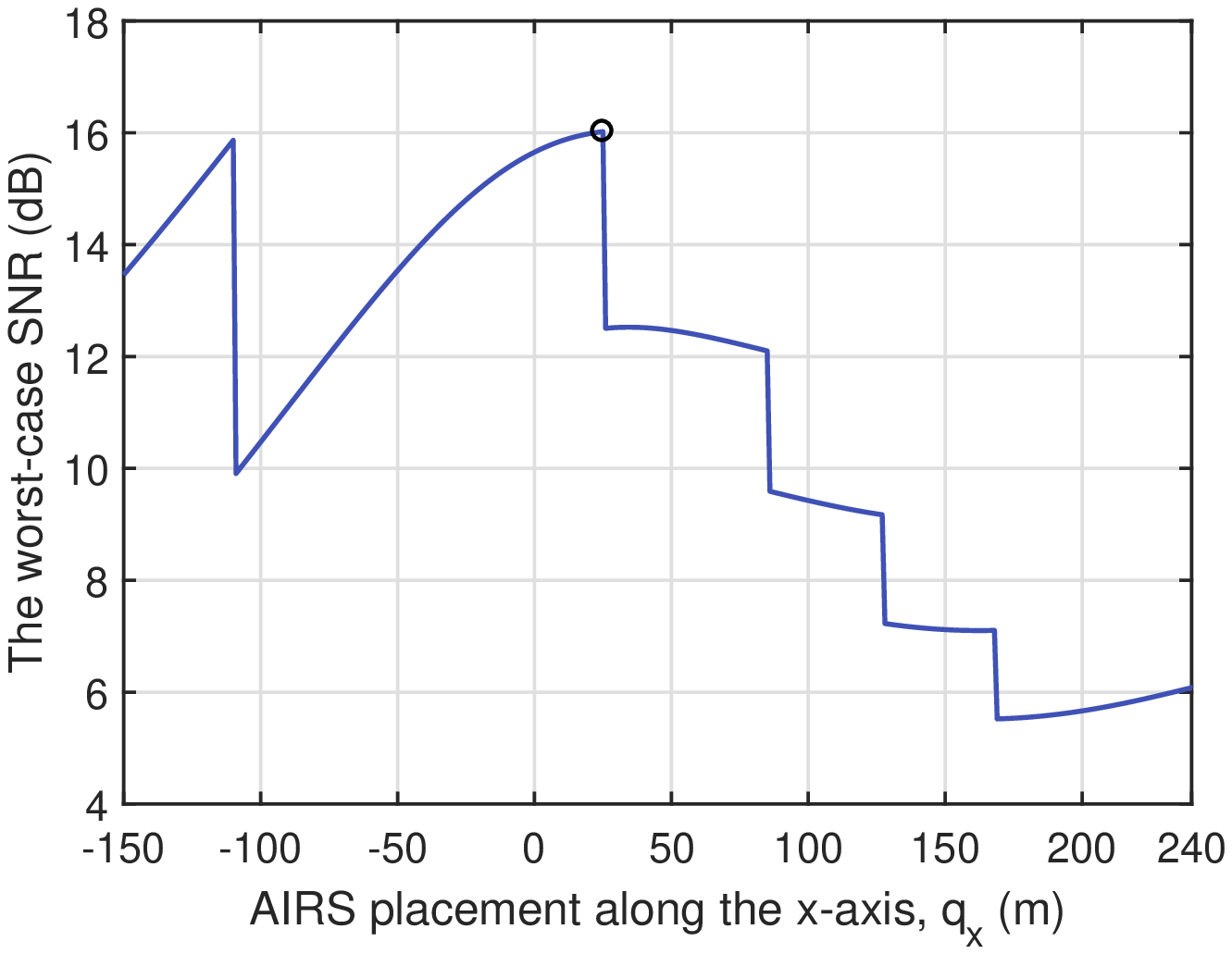}}
\vspace{0.3cm}
\end{minipage}
}
\caption{The worst-case SNR versus the AIRS placement along the $x$-axis for 1D coverage with the ULA-based AIRS.}
\label{SNRLineSegmentCase}
\end{figure*}

 Next, we consider the AIRS-enabled area coverage with the simplified ULA-based AIRS by assuming that $\mathcal A$ is also a one-dimensional (1D) line segment along the $x$-axis with interval $[x_l, x_u]$. Three different setups are considered, with $\left[ {{x_l},{x_u}} \right] = \left[ {250,750} \right]$~m, $\left[ {500,1500} \right]$~m, and $\left[ {155,325} \right]$~m, respectively. The number of reflecting elements is set as $N=N_x=256$. To illustrate the impact of AIRS placement $\bf q$ on each of the three factors in \eqref{AIRSPlacementOptimization3}, Fig.~\ref{LossandNumberLineSegmentCase} plots the required number of sub-arrays ${L^*}\left( q_x \right)$ in \eqref{theRequiredNumberofSubArrays} and the worst-case concatenated path loss versus the AIRS placement $q_x$ along the $x$-axis, together with the corresponding worst-case SNR plotted in Fig.~\ref{SNRLineSegmentCase}. Let ${x_c} = \frac{{{x_l} + {x_u}}}{2}$  be the center of the target line segment for AIRS coverage. It can be shown that the optimal AIRS placement should satisfy $q_x^* \le {x_c}$. Therefore, the maximum values over the $x$-axis in the plots of Fig.~\ref{LossandNumberLineSegmentCase} and Fig.~\ref{SNRLineSegmentCase} are set to $x_c$. It is firstly observed from Fig.~\ref{LossandNumberLineSegmentCase} that as the AIRS moves from the left of the source node towards $x_c$, the required number of sub-arrays ${L^*}\left( {{q_x}} \right)$ increases in a staircase manner. This is expected since when $q_x$ increases or the AIRS moves closer to the target coverage area, the angular span ${\Delta _{{\rm{span}}}}\left( {{q_x}} \right)$ associated with $\mathcal A$ increases, which thus requires broader beam (or equivalently more sub-array partitions) to cover the target area, although this comes at the cost of reduced worst-case array gain, as can be inferred from \eqref{beamBroadeningTechniqueArrayGain}. It is also observed from Fig.~\ref{LossandNumberLineSegmentCase} that the worst-case concatenated path loss generally first decreases, then increases and finally decreases with the increase of $q_x$.

 With the effects of both the angular span and concatenated path loss taken into account, the worst-case SNR versus $q_x$ based on \eqref{beamBroadeningTechniqueArrayGain} is shown in Fig.~\ref{SNRLineSegmentCase} for the three cases considered in Fig.~\ref{LossandNumberLineSegmentCase}. The optimal AIRS placement that leads to the best worst-case SNR is also labelled in the figure. It is observed that different from the single-location SNR maximization case, where the optimal AIRS placement is always located between the source node and the target location, the optimal AIRS placement $q_x^*$ for area coverage critically depends on the size of the target area, for which we might have $q_x^* < 0$, $q_x^* \approx 0$, or $0 < q_x^* \le {x_c}$, corresponding to the three cases in Fig.~\ref{SNRLineSegmentCase}, respectively. In particular, in Fig.~\ref{SNRLineSegmentCase}(a), the optimal AIRS lies to the left of the source node, since it can result in smaller angular span and hence higher worst-case array gain, though at the cost of higher concatenated path loss. This implies that the array gain is the dominating factor for the considered setup. Similar observations can be made for the two other cases in Fig.~\ref{SNRLineSegmentCase}. These three different setups show that it is a non-trivial task to find the optimal AIRS placement for area coverage in general, as it needs to achieve an optimal balance between minimizing the angular span and the concatenated path loss.

\begin{figure}[!t]
  \centering
  \centerline{\includegraphics[width=3.5in,height=2.625in]{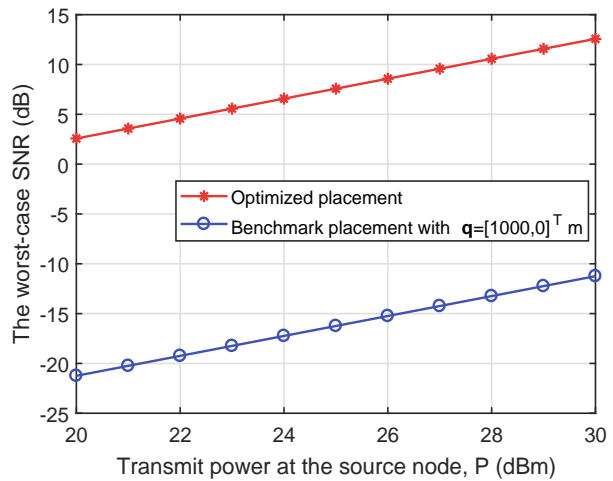}}
  \caption{The worst-case SNR versus transmit power for the ULA-based AIRS coverage.}
  \label{ULATheWorst-CaseSNRVersusTransmitPowerAtTheSourceNode}
\end{figure}
 Next, for the ULA-based AIRS, we consider the AIRS-enabled coverage for a rectangular area $\mathcal A$, with the length and width given by $D_x=1000$~m and $D_y=600$~m, respectively, and the center at ${{\bf{w}}_0} = {[1000,0]^T}$~m. The number of AIRS elements is set as $N=256$. For comparison, we consider the benchmark scheme with the AIRS placed above the center of the rectangular target area, i.e., ${\bf{q}}=[1000,0]^T$~m. Fig.~\ref{ULATheWorst-CaseSNRVersusTransmitPowerAtTheSourceNode} shows the worst-case SNR versus the transmit power $P$ for both the optimized and benchmark schemes. It is observed that the optimized AIRS placement significantly outperforms the benchmark placement, with an about 25~dB SNR gain. This demonstrates the importance of our proposed joint AIRS deployment and beamforming design.

 \begin{figure}[!t]
  \centering
  \centerline{\includegraphics[width=3.5in,height=2.625in]{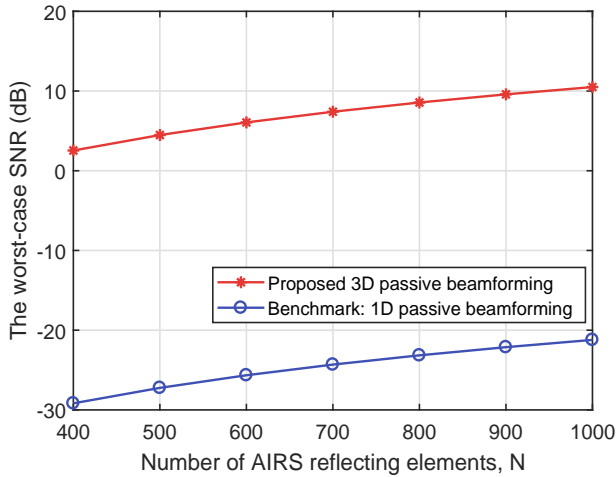}}
  \caption{The worst-case SNR versus the number of AIRS reflecting elements for the UPA-based AIRS coverage.}
  \label{UPATheWorst-CaseSNRVersusNumberofelements}
\end{figure}

  Last, we consider the general case of UPA-based AIRS as studied in Section~\ref{generalCaseofUPA-BasedAIRS}. The number of elements along the $y$-axis is fixed as $N_y=20$, and $N$ is varied by varying the number of elements along the $x$-axis, $N_x$. The size of the rectangular target area is the same as the ULA-based AIRS case considered previously. Besides the proposed scheme in Section~\ref{generalCaseofUPA-BasedAIRS}, we also consider a benchmark scheme with 1D passive beamforming applied to UPA-based AIRS, by grouping the $N_y$ reflecting elements in each column into one sub-surface and applying an identical phase shift~\cite{tang2019wireless}. In this case, the phase shift design of the AIRS with $N = {N_x}{N_y}$ elements follows the pattern ${\theta _{{n_x},{n_y}}} = {\theta _{{n_x}}},\forall {n_y}$. Such a design further restricts the optimization space in (P2), since the number of independent phase shift design variables is $N_x$, instead of $N_x+N_y$ in our proposed 3D beam design in Section~\ref{generalCaseofUPA-BasedAIRS}. By following the similar design of Proposition~\ref{Proposed Passive Beam Solution for P5.1}, the received SNR with the above 1D beamforming applied to UPA-based AIRS can be expressed as
 \begin{equation}
 \small
 \begin{aligned}
 &\gamma \left( {{\bf{q}},{\bf{\Theta }},{\bf{w}}} \right) \approx \\
 &\frac{4}{{{\pi ^2}}}\frac{{\bar P\beta _0^2MN_x^2{{\left| {\frac{{\sin \left( {\pi {N_y}{{\bar d}_y}\left[ {{{\bar \Omega }_T}\left( {{\bf{q}},{\bf{w}}} \right) - {{\bar \Omega }_R}\left( {\bf{q}} \right)} \right]} \right)}}{{\sin \left( {\pi {{\bar d}_y}\left[ {{{\bar \Omega }_T}\left( {{\bf{q}},{\bf{w}}} \right) - {{\bar \Omega }_R}\left( {\bf{q}} \right)} \right]} \right)}}} \right|}^2}}}{{{{\left\lceil {\sqrt {{\Delta _{{\rm{span}},x}}\left( {\bf{q}} \right){N_x}{{\bar d}_x}} } \right\rceil }^2}\left( {{H^2} + {{\left\| {{\bf{q}} - {\bf{w}}} \right\|}^2}} \right)\left( {{H^2} + {{\left\| {\bf{q}} \right\|}^2}} \right)}}. \label{BechmarkSchemeSNR}
  \end{aligned}
 \end{equation}
 In contrast to the broadened and flattened beam in the proposed 3D beamforming design, due to the inability of phase steering over the dimension of $\bar \Omega$, the SNR in \eqref{BechmarkSchemeSNR} varies significantly with the target location $\bf w$ and can be even zero when $\pi {N_y}{{\bar d}_y}\left[ {{{\bar \Omega }_T}\left( {{\bf{q}},{\bf{w}}} \right) - {{\bar \Omega }_R}\left( {\bf{q}} \right)} \right] = k\pi ,k = 1, \cdots ,{N_y} - 1$.

 Fig.~\ref{UPATheWorst-CaseSNRVersusNumberofelements} shows the worst-case SNR versus the number of AIRS reflecting elements $N$ for both the proposed and benchmark schemes. It is observed that the worst-case SNR increases with the number of AIRS reflecting elements for both schemes, as expected. Furthermore, the proposed 3D passive beamforming with beam broadening and flattening drastically outperforms the benchmark 1D passive beamforming applied to UPA-based AIRS, with an about 30~dB SNR gain. This is expected since the benchmark scheme is unable to achieve beam steering over $\bar \Omega$ dimension (see \eqref{BechmarkSchemeSNR}), and its resulting SNR is thus limited by the insufficient array gain along the $y$-axis, especially for a large coverage area. In contrast, thanks to the 3D beam broadening and flattening over both $\bar \Phi$ and $\bar \Omega$ dimensions, approximately equal array gain can be achieved for all locations in the target coverage area, thus leading to significant performance gain over the benchmark scheme.

 \begin{figure}[!t]
  \centering
  \centerline{\includegraphics[width=3.5in,height=2.625in]{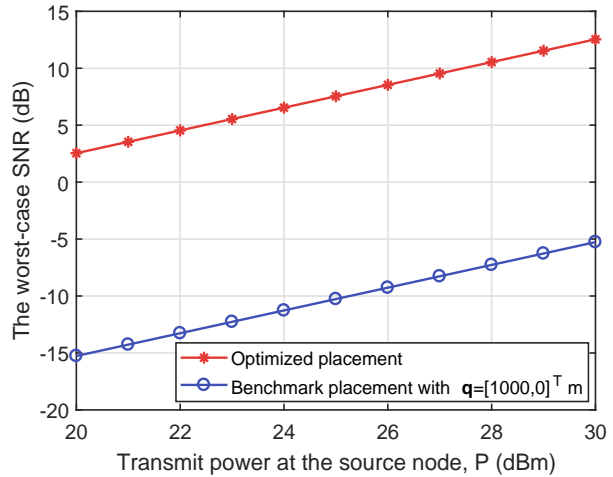}}
  \caption{The worst-case SNR versus transmit power for the UPA-based AIRS coverage.}
  \label{UPATheWorstCaseSNRVersusTransmitPowerAtTheSourceNode}
\end{figure}

Fig.~\ref{UPATheWorstCaseSNRVersusTransmitPowerAtTheSourceNode} shows the worst-case SNR versus transmit power at the source node for the UPA-based AIRS area coverage. The number of AIRS elements along $x$- and $y$-axis are set as $N_x=N_y=20$. We also consider the benchmark placement scheme with the AIRS placed above the center of the rectangular target area, i.e., ${\bf{q}}=[1000,0]^T$~m, and the proposed 3D beamforming in Section~\ref{generalCaseofUPA-BasedAIRS} is applied for both the optimized and benchmark placement schemes. It is observed that the optimized placement achieves significant performance gains over the benchmark placement. Again, this shows the importance of our proposed joint AIRS placement and beamforming design for AIRS-enabled wireless communications.

 \begin{figure}[!t]
  \centering
  \centerline{\includegraphics[width=3.5in,height=2.625in]{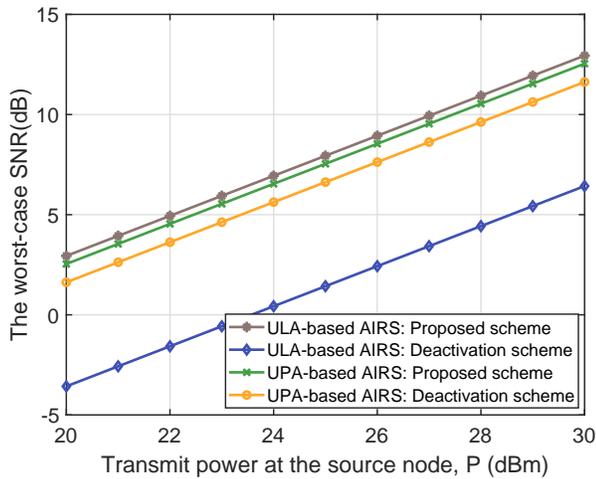}}
  \caption{ The worst-case SNR versus transmit power for the ULA- and UPA-based AIRS coverage.}
  \label{ULAUPAComparison}
\end{figure}
Fig.~\ref{ULAUPAComparison} shows the performance comparison between the ULA- and UPA-based AIRS coverage enhancement. The number of reflecting elements for both types of AIRS is set as $N=400$, where $N_x=N_y=20$ for the UPA-based AIRS. For comparison, the conventional deactivation-based beam broadening technique by turning off part of the reflecting elements is also considered~\cite{wang2009beam,xiao2016hierarchical}. It is firstly observed that with the proposed beam broadening and flattening technique, the ULA- and UPA-based AIRS achieve similar beamforming gain, and the proposed scheme outperforms the benchmark deactivation-based scheme. This is expected since with the proposed beam broadening and flattening technique, all locations in the target coverage area can achieve an approximately equal array gain. It is also observed that for the conventional deactivation-based scheme, the UPA-based AIRS obtains a better performance than the ULA-based AIRS. This is because the array aperture of the UPA is smaller than that of the ULA, and the required number of deactivation elements for the UPA to cover the target area is smaller than that for the ULA. Therefore, more beamforming gain is achieved for the UPA-based AIRS in the case of deactivation-based scheme.
\section{Conclusion}
This paper proposed a new wireless relaying system enabled by passive AIRS for coverage extension over the sky. The worst-case SNR in a target coverage area was maximized by jointly optimizing the transmit beamforming for the source node as well as the placement and 3D passive beamforming for the AIRS. We first studied the special case of single-location SNR maximization and derived the optimal solution in closed-form. It was shown that the optimal horizontal AIRS placement only depends on the ratio between the source-destination distance and the AIRS altitude. Then for the general case of AIRS-enabled area coverage, we proposed an efficient solution by decoupling the passive beamforming design for the AIRS to maximize the worst-case array gain from the AIRS placement optimization via balancing the resulting angular span and cascaded path loss. In particular, a novel beam broadening and flattening technique was applied to form a flattened beam pattern with broadened/adjustable beamwidth catering to the size of the target coverage area. Numerical results demonstrated that the proposed design can significantly improve the performance over the benchmark beamforming/placement schemes, and the importance of exploiting both the AIRS passive beamforming and flexible deployment in designing AIRS-enabled wireless communications.

\bibliographystyle{IEEEtran}
\bibliography{ref}

\begin{thebibliography}{10}
\providecommand{\url}[1]{#1}
\csname url@samestyle\endcsname
\providecommand{\newblock}{\relax}
\providecommand{\bibinfo}[2]{#2}
\providecommand{\BIBentrySTDinterwordspacing}{\spaceskip=0pt\relax}
\providecommand{\BIBentryALTinterwordstretchfactor}{4}
\providecommand{\BIBentryALTinterwordspacing}{\spaceskip=\fontdimen2\font plus
\BIBentryALTinterwordstretchfactor\fontdimen3\font minus
  \fontdimen4\font\relax}
\providecommand{\BIBforeignlanguage}[2]{{%
\expandafter\ifx\csname l@#1\endcsname\relax
\typeout{** WARNING: IEEEtran.bst: No hyphenation pattern has been}%
\typeout{** loaded for the language `#1'. Using the pattern for}%
\typeout{** the default language instead.}%
\else
\language=\csname l@#1\endcsname
\fi
#2}}
\providecommand{\BIBdecl}{\relax}
\BIBdecl

\bibitem{lu2020enabling}
H.~Lu, Y.~Zeng, S.~Jin, and R.~Zhang, ``Enabling panoramic full-angle
  reflection via aerial intelligent reflecting surface,'' in \emph{Proc. IEEE
  ICC Workshop}, Jun. 2020.

\bibitem{Latvaaho2019Keydrivers}
M.~Latva-aho and K.~Lepp{\"a}nen, ``Key drivers and research challenges for
  6{G} ubiquitous wireless intelligence,'' 6G Flagship, University of Oulu,
  Finland, Sep. 2019.

\bibitem{letaief2019roadmap}
K.~B. Letaief, W.~Chen, Y.~Shi, J.~Zhang, and Y.~A. Zhang, ``The roadmap to
  6{G}: A{I} empowered wireless networks,'' \emph{IEEE Commun. Mag.}, vol.~57,
  no.~8, pp. 84--90, Aug. 2019.

\bibitem{yang20196g}
P.~Yang, Y.~Xiao, M.~Xiao, and S.~Li, ``6{G} wireless communications: Vision
  and potential techniques,'' \emph{IEEE Netw.}, vol.~33, no.~4, pp. 70--75,
  Jul./Aug. 2019.

\bibitem{wang2009beam}
J.~Wang \emph{et~al.}, ``Beam codebook based beamforming protocol for
  multi-{G}bps millimeter-wave {WPAN} systems,'' \emph{IEEE J. Sel. Areas
  Commun.}, vol.~27, no.~8, pp. 1390--1399, Oct. 2009.

\bibitem{zhang2005variable-phase-shift-based}
X.~Zhang, A.~F. Molisch, and S.-Y. Kung, ``Variable-phase-shift-based
  {RF}-baseband codesign for {MIMO} antenna selection,'' \emph{IEEE Trans.
  Signal Process.}, vol.~53, no.~11, pp. 4091--4103, Nov. 2005.

\bibitem{ayach2014spatially}
O.~E. Ayach, S.~Rajagopal, S.~Abu-Surra, Z.~Pi, and R.~W. Heath, Jr.,
  ``Spatially sparse precoding in millimeter wave {MIMO} systems,'' \emph{IEEE
  Trans. Wireless Commun.}, vol.~13, no.~3, pp. 1499--1513, Mar. 2014.

\bibitem{zeng2016millimeter}
Y.~Zeng and R.~Zhang, ``Millimeter wave {MIMO} with lens antenna array: A new
  path division multiplexing paradigm,'' \emph{IEEE Trans. Commun.}, vol.~64,
  no.~4, pp. 1557--1571, Apr. 2016.

\bibitem{singh2009on}
J.~Singh, O.~Dabeer, and U.~Madhow, ``On the limits of communication with
  low-precision analog-to-digital conversion at the receiver,'' \emph{IEEE
  Trans. Commun.}, vol.~57, no.~12, pp. 3629--3639, Dec. 2009.

\bibitem{zhang2018on}
J.~Zhang, L.~Dai, X.~Li, Y.~Liu, and L.~Hanzo, ``On low-resolution {ADC}s in
  practical 5{G} millimeter-wave massive {MIMO} systems,'' \emph{IEEE Commun.
  Mag.}, vol.~56, no.~7, pp. 205--211, Jul. 2018.

\bibitem{subrt2012intelligent}
L.~Subrt and P.~Pechac, ``Intelligent walls as autonomous parts of smart indoor
  environments,'' \emph{IET Commun.}, vol.~6, no.~8, pp. 1004--1010, May 2012.

\bibitem{tan2016increasing}
X.~Tan, Z.~Sun, J.~M. Jornet, and D.~Pados, ``Increasing indoor spectrum
  sharing capacity using smart reflect-array,'' in \emph{Proc. IEEE ICC}, May
  2016, pp. 1--6.

\bibitem{tang2019wireless}
W.~Tang, X.~Li, J.~Y. Dai, S.~Jin, Y.~Zeng, Q.~Cheng, and T.~J. Cui, ``Wireless
  communications with programmable metasurface: Transceiver design and
  experimental results,'' \emph{China Commun.}, vol.~16, no.~5, pp. 46--61, May
  2019.

\bibitem{di2019smart}
M.~Di~Renzo \emph{et~al.}, ``Smart radio environments empowered by
  reconfigurable {AI} meta-surfaces: An idea whose time has come,''
  \emph{EURASIP J. Wireless Commun. Netw.}, vol. 2019, no.~1, pp. 1--20, May
  2019.

\bibitem{han2019large}
Y.~Han, W.~Tang, S.~Jin, C.~Wen, and X.~Ma, ``Large intelligent
  surface-assisted wireless communication exploiting statistical {CSI},''
  \emph{IEEE Trans. Veh. Technol.}, vol.~68, no.~8, pp. 8238--8242, Aug. 2019.

\bibitem{huang2019reconfigurable}
C.~Huang, A.~Zappone, G.~C. Alexandropoulos, M.~Debbah, and C.~Yuen,
  ``Reconfigurable intelligent surfaces for energy efficiency in wireless
  communication,'' \emph{IEEE Trans. Wireless Commun.}, vol.~18, no.~8, pp.
  4157--4170, Aug. 2019.

\bibitem{basar2019wireless}
E.~Basar, M.~Di~Renzo, J.~De~Rosny, M.~Debbah, M.~S. Alouini, and R.~Zhang,
  ``Wireless communications through reconfigurable intelligent surfaces,''
  \emph{IEEE Access}, vol.~7, pp. 116\,753--116\,773, Aug. 2019.

\bibitem{cui2019secure}
M.~Cui, G.~Zhang, and R.~Zhang, ``Secure wireless communication via intelligent
  reflecting surface,'' \emph{IEEE Wireless Commun. Lett.}, vol.~8, no.~5, pp.
  1410--1414, Oct. 2019.

\bibitem{wu2019intelligent}
Q.~Wu and R.~Zhang, ``Intelligent reflecting surface enhanced wireless network
  via joint active and passive beamforming,'' \emph{IEEE Trans. Wireless
  Commun.}, vol.~18, no.~11, pp. 5394--5409, Nov. 2019.

\bibitem{wu2019towards}
------, ``Towards smart and reconfigurable environment: Intelligent reflecting
  surface aided wireless network,'' \emph{IEEE Commun. Mag.}, vol.~58, no.~1,
  pp. 106--112, Jan. 2020.

\bibitem{bjornson2020intelligent}
E.~Bj{\"o}rnson, {\"O}.~{\"O}zdogan, and E.~G. Larsson, ``Intelligent
  reflecting surface vs. decode-and-forward: How large surfaces are needed to
  beat relaying?'' \emph{IEEE Wireless Commun. Lett.}, vol.~9, no.~2, pp.
  244--248, Feb. 2020.

\bibitem{Yang2020IntelligentRS}
Y.~Yang, B.~Zheng, S.~Zhang, and R.~Zhang, ``Intelligent reflecting surface
  meets {OFDM}: Protocol design and rate maximization,'' \emph{IEEE Trans.
  Commun.}, vol.~68, no.~7, pp. 4522--4535, Jul. 2020.

\bibitem{Tang2019WirelessCW}
W.~Tang, M.~Z. Chen, X.~Chen, J.~Y. Dai, Y.~Han, M.~D. Renzo, Y.~Zeng, S.~Jin,
  Q.~Cheng, and T.~J. Cui, ``Wireless communications with reconfigurable
  intelligent surface: Path loss modeling and experimental measurement,''
  \emph{IEEE Trans. Wireless Commun.}, 2020, {D}OI: 10.1109/TWC.2020.3024887.

\bibitem{wu2020intelligentTutorial}
Q.~Wu, S.~Zhang, B.~Zheng, C.~You, and R.~Zhang, ``Intelligent reflecting
  surface aided wireless communications: A tutorial,'' 2020, arXiv:2007.02759.
  [Online]. {A}vailable: \url{https://arxiv.org/abs/2007.02759}.

\bibitem{huang2020achievable}
W.~Huang, Y.~Zeng, and Y.~Huang, ``Achievable rate region of {MISO}
  interference channel aided by intelligent reflecting surface,'' to appear in
  \emph{IEEE Trans. Veh. Technol.}, {A}vailable:
  \url{https://arxiv.org/abs/2005.09197}.

\bibitem{huang2020holographic}
C.~Huang \emph{et~al.}, ``Holographic {MIMO} surfaces for 6{G} wireless
  networks: Opportunities, challenges, and trends,'' \emph{IEEE Wireless
  Commun.}, vol.~27, no.~5, pp. 118--125, Oct. 2020.

\bibitem{alexandropoulos2020reconfigurable}
G.~C. Alexandropoulos, G.~Lerosey, M.~Debbah, and M.~Fink, ``Reconfigurable
  intelligent surfaces and metamaterials: The potential of wave propagation
  control for 6{G} wireless communications,'' \emph{IEEE ComSoc TCCN
  Newslett.}, vol.~6, no.~1, pp. 25--37, Jun. 2020.

\bibitem{shlezinger2020dynamic}
N.~Shlezinger, G.~C. Alexandropoulos, M.~F. Imani, Y.~C. Eldar, and D.~R.
  Smith, ``Dynamic metasurface antennas for 6{G} extreme massive {MIMO}
  communications,'' 2020, arXiv:2006.07838. [Online]. {A}vailable:
  \url{https://arxiv.org/abs/2006.07838}.

\bibitem{zeng2019accessing}
Y.~Zeng, Q.~Wu, and R.~Zhang, ``Accessing from the sky: A tutorial on {UAV}
  communications for 5{G} and beyond,'' \emph{Proc. IEEE}, vol. 107, no.~12,
  pp. 2327--2375, Dec. 2019.

\bibitem{hur2013millimeter}
S.~Hur, T.~Kim, D.~J. Love, J.~V. Krogmeier, T.~A. Thomas, and A.~Ghosh,
  ``Millimeter wave beamforming for wireless backhaul and access in small cell
  networks,'' \emph{IEEE Trans. Commun.}, vol.~61, no.~10, pp. 4391--4403, Oct.
  2013.

\bibitem{zhang2017codebook}
J.~Zhang, Y.~Huang, Q.~Shi, J.~Wang, and L.~Yang, ``Codebook design for beam
  alignment in millimeter wave communication systems,'' \emph{IEEE Trans.
  Commun.}, vol.~65, no.~11, pp. 4980--4995, Nov. 2017.

\bibitem{zhu20193D}
L.~Zhu, J.~Zhang, Z.~Xiao, X.~Cao, D.~O. Wu, and X.~Xia, ``3-{D} beamforming
  for flexible coverage in millimeter-wave {UAV} communications,'' \emph{IEEE
  Wireless Commun. Lett.}, vol.~8, no.~3, pp. 837--840, Jun. 2019.

\bibitem{zhou2020framework}
G.~Zhou, C.~Pan, H.~Ren, K.~Wang, and A.~Nallanathan, ``A framework of robust
  transmission design for {IRS}-aided {MISO} communications with imperfect
  cascaded channels,'' \emph{IEEE Trans. Signal Process.}, vol.~68, pp.
  5092--5106, Aug. 2020.

\bibitem{zhou2020user}
G.~Zhou, C.~Pan, H.~Ren, K.~Wang, A.~Nallanathan, and K.-K. Wong, ``User
  cooperation for {IRS}-aided secure {SWIPT} {MIMO}: Active attacks and passive
  eavesdropping,'' 2020, arXiv:2006.05347. [Online]. {A}vailable:
  \url{https://arxiv.org/abs/2006.05347}.

\bibitem{wei2020channel}
L.~Wei, C.~Huang, G.~C. Alexandropoulos, C.~Yuen, Z.~Zhang, and M.~Debbah,
  ``Channel estimation for {RIS}-empowered multi-user {MISO} wireless
  communications,'' 2020, arXiv:2008.01459. [Online]. {A}vailable:
  \url{https://arxiv.org/abs/2008.01459}.

\bibitem{han2013relay}
L.~Han, C.~Huang, S.~Shao, and Y.~Tang, ``Relay placement for
  amplify-and-forward relay channels with correlated shadowing,'' \emph{IEEE
  Wireless Commun. Lett.}, vol.~2, no.~2, pp. 171--174, Apr. 2013.

\bibitem{mailloux2005phased}
R.~J. Mailloux, \emph{Phased array antenna handbook}.\hskip 1em plus 0.5em
  minus 0.4em\relax Artech house, 2005.

\bibitem{rajagopal2012beam}
S.~Rajagopal, ``Beam broadening for phased antenna arrays using multi-beam
  subarrays,'' in \emph{Proc. IEEE Int. Conf. Commun. (ICC)}, Jun. 2012, pp.
  3637--3642.

\bibitem{Orfanidis2016Electromagnetic}
S.~J. Orfanidis, ``Electromagnetic waves and antennas,'' 2016, [Online].
  {A}vailable: \url{http://eceweb1.rutgers.edu/\%7eorfanidi/ewa/}.

\bibitem{xiao2016hierarchical}
Z.~Xiao, T.~He, P.~Xia, and X.-G. Xia, ``Hierarchical codebook design for
  beamforming training in millimeter-wave communication,'' \emph{IEEE Trans.
  Wireless Commun.}, vol.~15, no.~5, pp. 3380--3392, May 2016.

\end{thebibliography}
\end{document}